\documentclass[12pt]{article} 
\usepackage{amssymb}
\usepackage{graphicx}
\begin{document} 

\begin{center}
          {\Large \bf Elastic scattering of hadrons} 

\vspace{0.5cm}                   
{\bf I.M. Dremin\footnote{Email: dremin@lpi.ru}}

\vspace{0.5cm}              
          {\it Lebedev Physical Institute, Moscow 119991, Russia}

\vspace{5mm}

\hfill If only you knew what trash gives rise\\
\hfill To verses that aren't ashamed uprise \\
\vspace{2mm}
\hfill A. Akhmatova \\

\vspace{5mm}
{\bf \Large Contents}

\end{center}
$\;\;\;\;$ {\bf \large 1 Introduction.}

\medskip

{\bf \large 2 The main relations.}

\medskip

{\bf \large 3 Where do we stand now?}

\medskip

{\bf \large 4 Experimental data and phenomenological models.}
        
{\bf \large 4.1 The diffraction cone and geometrical approach. }

{\bf 4.1.1 The geometry of the internal hadron structure.}

{\bf 4.1.2 The modified Fraunhofer diffraction.}

{\bf 4.1.3 Electromagnetic analogies.}

{\bf 4.1.4 Reggeon exchanges.}

{\bf 4.1.5 The QCD-inspired models.}

{\bf \large 4.2 The intermediate angles: the dip and the Orear regime.}

{\bf 4.2.1 The Gaussian fits.}

{\bf 4.2.2 The phenomenological models.}

{\bf 4.2.3 The unitarity condition.}

{\bf 4.2.4 The overlap function and the eikonal.}

{\bf 4.2.5 The real part of the elastic scattering amplitude at nonzero 
transferred momenta.}

{\bf \large 4.3 Scaling laws.}

{\bf \large 4.4 The hard scattering at large angles. }

{\bf 4.4.1 The dimensional counting. }

{\bf 4.4.2 Coherent scattering.}

\medskip

{\bf \large 5 Discussion and conclusions.}

\medskip

{\bf \large References.}

\vspace{2mm}



\begin{abstract}
Colliding high energy hadrons either produce new particles or
scatter elastically with their quantum numbers conserved and no other particles 
produced.  We consider the latter case here. 
Although inelastic processes dominate at high energies, elastic scattering 
contributes considerably (18-25$\% $) to the total cross section. 
Its share first decreases and then increases at higher energies. 
Small-angle scattering prevails at all energies. Some characteristic features
are seen that provide informationon the geometrical structure of the colliding 
particles and the relevant dynamical mechanisms. The steep Gaussian peak 
at small angles is followed by the exponential (Orear) regime with some 
shoulders and dips, and then by a power-law drop. 

Results from various theoretical approaches are compared with experimental data. 
Phenomenological models claiming to describe this process are reviewed.
The unitarity condition predicts an exponential fall for the differential
cross section with an additional substructure to occur exactly between the 
low momentum transfer diffraction cone  
and a power-law, hard parton scattering regime under high momentum transfer. 
Data on the interference of the Coulomb and nuclear parts of amplitudes at 
extremely small angles provide the value of the real part of the forward 
scattering nuclear amplitude. 

The real part of the elastic scattering amplitude and 
the contribution of inelastic processes to the imaginary part of this amplitude 
(the so-called overlap function) at nonforward transferred momenta are also 
discussed. Problems related to the scaling behavior of the differential
cross section are considered.
The power-law regime at highest momentum transfer is briefly described.
\end{abstract}

\section{Introduction}

Hadron interactions are strong and, in principle, should be described
by quantum chromodynamics (QCD). However, experimental data show that their 
main features originate from the non-perturbative sector of QCD. Only  
comparatively rare processes with large transferred momenta can be treated 
theoretically rather successfully by perturbative methods due to the 
well-known property of the asymptotical freedom of QCD. Hence, in the absence of
methods for a rigorous solution of QCD equations, our understanding 
of the dynamics of the main bulk of strong interactions is severely limited 
by model building or some rare rigorous relations. In fact, our approach 
to high-energy hadronic processes at present is at best still in its infancy.

As has been learned from experiment, strong interactions of colliding 
high-energy particles give rise to inelastic and elastic processes. Some new 
particles (mostly pions) are produced in inelastic processes, which are the most 
probable ones, comprising 75$\% $ to 80$\% $ of all processes at high energies. 
 Most created particles have comparatively small transverse momenta.

At the same time, in 25$\% $ to 20$\% $ of events, the colliding particles 
do not change their nature and scatter elastically, declining at some 
angle from their initial trajectories. The only information about this process
available from experiment is obtained by the measurement of the differential 
cross section (proportional to the probability) of elastic scattering at some
angle at a given energy.

In a very tiny range of extremely
small angles, the charged particles scatter due to electromagnetic forces.
But the dominant process of elastic scattering due to strong
interactions proceeds at somewhat larger angles in the so-called diffraction 
cone. The differential cross sections are heavily weighted toward small 
transferred  momenta exhibiting a huge peak. The scattering angle is still 
rather small there and becomes smaller and smaller as the energy increases. The 
probability of scattering at a given angle in this region 
decreases steeply, similarly to a Gaussian exponential. Noticeably 
less than one percent of particles are elastically scattered to larger angles 
outside this diffraction cone. The Gaussian behavior is replaced there by 
a simple exponential one with some shoulders and (or) dips. At ever
larger angles (or transferred momenta), a power-like decrease has been 
observed. At angles close to $\pi /2$, some additional flattening is seen.

The elastic cross section (the integral of the differential distribution over
angles or transverse momentum) depends on the energy of the colliding partners. 
At high energies, it shows a steady tendency to become larger with an 
increase in energy. We note that the inelastic cross section also increases,
such that their sum (the total cross section) increases as well.   

The process of elastic scattering of hadrons has been studied experimentally in 
a wide energy region with different initial particles. At high energies of 
colliding partners, the most detailed results are available for the scattering 
of protons ($pp$) and antiprotons ($p\bar p$) on protons. We mainly 
discuss these data, sometimes referring to other colliding partners 
of protons such as pions and kaons.

Some surprises in the behavior of the differential cross sections
appeared in the 1960s when the very first experimental data on 
elastic $pp$ and $\pi p$ scattering were obtained at energies between 6.8 and 
19.2 GeV in the laboratory system \cite{fole, cocc, krisc, nara, orea, bake,
cocc1, orear, alla} (the total energy in the center-of-mass system (cms) is only 
$\sqrt s \approx $ 4 - 6 GeV !). The diffraction cone behavior changed at 
larger transferred momenta $\vert t\vert $ to a slower $t$-dependence. Somewhat
later, the energy range was extended to 50 GeV \cite{aker, ayre, asad}. With 
the advent of new accelerators, the data for $pp$ scattering at energies 
$\sqrt s\approx $ 19, 20, 23, 28, 31, 45, 53, 62 GeV were published \cite{ama, ame, 
bohm, kwak, alb, hart, cone, baks, nagy, amal, kuzn, schiz, fide, fais, rubi, bre} 
and the data for $p\bar p$ at 31, 53, 62, 546, 630, 1800, 1980 GeV 
\cite{eise, amo, ambr, saka, boz, bre1, ber, aug, abe, teva} appeared. 
The early results are reviewed in Refs \cite{zrt, uama}.
The compilation of the data can be found in \cite{cude}. Only recently, the 
results of the TOTEM collaboration at the LHC on elastic $pp$ scattering 
processes at $\sqrt s$ = 7 TeV were published \cite{toteml, totemh}. 

Surely, these results asked for their understanding and theoretical 
interpretation. The most important task is to acquire some knowledge about the
internal structure of colliding particles by deciphering the information
supplied by experimental data about the dependence on energy and transferred
momentum. The transferred momentum is directly related to the size and the
structure of those regions inside the hadron that participate in the
interaction.

Many phenomenological models have been proposed. Most of them aspire to be 
"a phenomenology of 
everything"$\;$ related to elastic scattering of hadrons in a wide energy range. 
Doing so in the absence of applicable laws and methods of the fundamental 
theory, they have to use a large number of adjustable parameters. The free 
parameters have been determined by fitting the model results to the available
experimental data. Even then, their predictions often fail when a 
new energy domain becomes available. And "the verse"$\;$ does not grow anymore! 
(If not recultivated.) Independent of their success and failure, we are sure 
that, "in the long run, the physical picture may be expected to be much more 
important than most of the detailed computations" \cite{hctt}. In what follows, 
we mention and discuss many of them. 

The scattering of charged particles at extremely small angles is completely 
dominated by the Coulomb amplitude. The absolute value of the Born amplitude 
is well known. The phase of the Coulomb amplitude varies depending on the
model chosen. However, this variation is rather mild in the considered 
tiny region of extremely small angles. The interference of the Coulomb amplitude 
with the strong-interaction (nuclear) amplitude in the transition region where 
they are almost equal has been used for the experimental 
determination of the ratio of the real to imaginary parts of the latter.
This interference also depends on the chosen form of the nuclear amplitude. 
Theoretically, this ratio can be estimated with the help of dispersion
relations. We briefly discuss this problem and show how the 
obtained results influence our analysis of scattering at somewhat larger 
angles. 

The most numerous group of models deals with phenomenological attempts to 
describe the main bulk of elastic scattering at small angles in the diffraction 
cone. In general, they are based on some geometrical models of particle 
substructure, with peripheral regions playing the decisive role. The 
approach using the Reggeon (Pomeron) exchanges is the most popular among 
them. The approximately Gaussian (in angles) shape of the experimentally 
measured differential cross section in this region has been fitted just in 
this way. In addition to it, the simplest classical expressions for diffractive 
processes and results on the electromagnetic form factors are also used. 
However, the bold extension of the obtained results to larger angles is usually 
not very successful, even though some new parameters are introduced. 

Particles scattered at larger angles give insight into the deeper internal 
regions of particle structure. The multiple iteration (rescattering) of 
diffractive processes may
explain the region of angles that are somewhat larger than the diffractive ones. 
Without any additional model building, it can be described as a consequence of 
the unitarity condition. The only necessary input is the experimentally known 
energy behavior of the diffraction cone slope and the total cross section. 
It predicts the observed exponential fall-off with angles and 
damped oscillations imposed on it, which, depending on their amplitudes, lead 
to shoulders or dips of the differential cross sections. 

At somewhat larger angles, the elastic processes may be considered to be 
dominated by the innermost constituents of the colliding particles. The 
perturbative QCD approach to hard parton scattering convoluted with some results 
on the parton structure of colliding particles is then used to describe 
experimental data. This approach predicts the power-like angular dependence of 
the differential cross sections. It has been seen in experiment. The dimensional 
(or quark) counting of the number of participating partons has been successful.
The convolution with the internal structure of particles implies some coherence 
in the behavior of its constituents: all of them should coherently turn at the 
same angle. A particle should not be destroyed during the collision, and its 
internal wave function must be left intact. Therefore we can call such 
processes coherent large-angle scattering.   

At angles close to $\pi /2$, the effects of symmetrization of the
corresponding amplitudes may become important and lead to some flattening of 
the differential distribution.

There are no strict definitions of the lower and upper bounds of these regions.
The diffraction peak shrinks with energy, such that the exponential fall-off 
with squared transferred momenta $t$ terminates at ever smaller values. 
Correspondingly, the dip after it shifts to smaller values of $\vert t\vert $ 
as does the $\sqrt {\vert t\vert }$-exponential. At low energies, this regime
approximately occupies the interval between 0.8 and 2 GeV$^2$, while in the LHC,
 it has moved to 0.4 - 1.5 GeV$^2$. According to the QCD prejudice, the scale 
for parton scattering should be set above 1 GeV$^2$. This is actually observed
with a power-like decrease starting somewhere around $\vert t\vert >$1.5 - 2 
GeV$^2$ at the LHC.

Hence, we can speak, at least, about five subregions of elastic scattering.
We mainly discuss three of them: the diffraction cone, the Orear regime
and coherent hard parton scattering. The diffraction cone is well known to us
from semiclassical effects. The regions beyond it became noticeable only at 
energies of colliding particles above several GeV, where processes of scattering 
at sufficiently large angles or transferred momenta 
are observable. They persist up to the present LHC energy of 7 TeV. Who ordered 
them and whether they will they survive at ever higher energies are also the 
questions to be discussed in this review.

Its structure of this paper is as follows. The main relations between different 
characteristics of elastic scattering
are presented in Section 2. Then, in Section 3, their global
dependences on energy and transferred momenta are discussed, together with
our attempts to understand their implications within the simplest approaches.
A more detailed analysis of experimental data in the framework of different
theoretical ideas and approximations is the content of Section 4. Finally,
the general picture is briefly discussed in Section 5.

We do not consider the scattering of polarized particles, and the spin structure
of the amplitude is ignored.

\section{The main relations}

As discussed above, the measurement of the differential cross section is
the only source of experimental information about a process. Hence,
the main characteristics of hadron interactions directly related to the elastic 
scattering amplitude, such as the total cross section, the elastic scattering
cross section, the ratio of the real to the imaginary part of the amplitude, 
and the slope of the diffraction cone, are obtained. The first two 
are functions of the total energy only, while the others depend on two variables:
the total energy and the transferred momentum (or the scattering angle).

The dimensionless elastic scattering amplitude $A$ defines the 
differential cross section as
\begin{equation}
\frac {d\sigma (s)}{dt}=\frac {1}{16\pi s^2}\vert A\vert ^2=
\frac {1}{16\pi s^2}({\rm Im}A(s,t))^2(1+\rho ^2(s,t)),
\label{dsel}
\end{equation}
where the ratio of the real to imaginary parts of the amplitude is 
defined:
\begin{equation}
\rho (s,t) = \frac {{\rm Re}A(s,t) }{{\rm Im}A(s,t)}.
\label{rho}
\end{equation}
In what follows, we consider very high energy processes. Therefore,
the masses of the colliding particles can be neglected, and we use the 
expression $s=4E^2\approx 4p^2$, where $E$ and $p$ are the energy and the
momentum in the center-of-mass system. The four-momentum transfer squared is
\begin{equation}
-t=2p^2(1-\cos \theta )\approx p^2\theta ^2\approx p_t^2 \;\;\;\;\; 
(\theta \ll 1)
\label{trans}
\end{equation}
with $\theta $ denoting the scattering angle in the center-of-mass system
and $p_t$ being the transverse momentum. 

The elastic scattering cross section is given by the integral of the
differential cross section (\ref{dsel}) over all transferred 
momenta:
\begin{equation}
\sigma _{el}(s)=\int_{t_{min}}^{0} dt \frac {d\sigma (s)}{dt}.
\label{sel}
\end{equation}
              
The total cross section $\sigma _t$ is related by the optical theorem to
the imaginary part of the forward scattering amplitude as
\begin{equation}
\sigma _t(s)= \frac {{\rm Im}A(p,\theta =0) }{s}.
\label{sigt}
\end{equation}     

Elastically scattered hadrons escape from the interaction region declining
mostly at quite small angles within the so-called diffraction cone\footnote {In
practice, the tiny region of the interference of the 
Coulomb and nuclear amplitudes at extremely small angles does not contribute 
to the total cross section of elastic scattering. Its role in
obtaining some estimates of $\rho (s, t=0)$ is described below.}. Therefore 
the main focus has been on this 
region. As known from experiment, the diffraction peak has a Gaussian shape in 
the scattering angles or exponentially decreases as a function of the 
transferred momentum squared:
\begin{equation}\
\frac {d\sigma }{dt}/\left( \frac {d\sigma }{dt}\right )_{t=0}=e^{Bt}\approx 
e^{-Bp^2\theta ^2}.
\label{diff}
\end{equation}
In view of relations (\ref{sel}), (\ref{sigt}), (\ref{diff}), any successful 
theoretical description of the differential distribution must also work in 
fitting the energy dependences of the total and elastic cross sections.
 
The diffraction cone slope $B$ is given by
\begin{equation}
B(s,t)\approx \frac {d}{dt} \left [\ln \frac {d\sigma (s,t)}{dt}\right ].
\label{bst}
\end{equation}
Actually, the slope $B$ depends slightly on $t$ at the given energy $s$, 
e.g., at the LHC, its value changes by about 
10$\%$ within the cone for $\vert \Delta t\vert \approx 0.3$ GeV$^2$. We neglect
this in the first approximation.

The normalization factor in Eq. (\ref{diff}) is
\begin{equation}
 \left (\frac {d\sigma }{dt}\right )_{t=0}=\frac {\sigma _t^2(s)(1+\rho _0^2(s))}
{16\pi }, 
\label{diff0}
\end{equation}
where $\rho _0$ is defined as the value of the ratio of the real and imaginary 
parts of the amplitude in the forward direction at $\theta = t$ = 0.
Eq. (\ref{diff0}) follows from formula (\ref{dsel}) and optical theorem 
(\ref{sigt}) at $t=0$.

According to the dispersion relations, which connect the real and imaginary parts 
of the amplitude, and optical theorem Eq. (\ref{sigt}), the value $\rho _0$ 
can be expressed as an integral
of the total cross section over the whole energy range. In practice, $\rho _0$
is mainly sensitive to the local derivative of the total cross section. 
In the first approximation, the result of the dispersion relation can then 
be written in the form \cite{gmig, sukha, fkol}
\begin{equation}
\rho _0(s)\approx \frac {1}{\sigma _t}\left [\tan \left (\frac {\pi }{2}
\frac {d}{d\ln s }\right )\right ]\sigma _t=
\frac {1}{\sigma _t}\left [ \frac {\pi }{2}\frac {d}{d\ln s }+
\frac {1}{3}\left (\frac {\pi }{2}\right )^3\frac {d^3}{d\ln s^3 }+...\right ]
\sigma _t.
\label{rhodi}
\end{equation}
It follows that at high energies $\rho _0(s)$ is mainly determined by the 
derivative of the logarithm of the total cross section with respect to the 
logarithm of energy.

The bold extension of the first term in this series to nonzero transferred
momenta would look like
\begin{equation}
\rho (s,t)\approx \frac {\pi }{2}
\left [\frac {d\ln {\rm Im}A(s,t) }{d\ln s }-1\right ].
\label{rhodit}
\end{equation}

If we neglect the high-$\vert t\vert $ tail of the differential cross section,
which is several orders of magnitude lower than the optical point, and 
integrate in Eq. (\ref{sel}) using  expression (\ref{diff}) with constant $B$, 
we obtain the approximate relation between the total cross section, the elastic 
cross section, and the slope: 
\begin{equation}
\frac {\sigma _t^2(1+\rho _0^2)}{16\pi B\sigma _{el}}\approx 1.
\label{stseb}
\end{equation}

We can compare this formula with the upper bound obtained in Ref. \cite{mcdm}:
\begin{equation}
\frac {\sigma _t^2}{18\pi B\sigma _{el}}\leq 1.
\label{mcdm}
\end{equation}

The phase $\zeta $ of the hadronic amplitude is often defined as
\begin{equation}
           A(s,t)=i\vert A(s,t)\vert e^{-i\zeta (s,t)};
\label{zeta}
\end{equation}
then
\begin{equation}
      \rho (s,t)=\tan \zeta (s,t).
\label{rzet}
\end{equation}

These formulas are used for measuring the luminosity, which relates
the cross section $\sigma _i$ of a given process $i$ to the corresponding
number of events $N_i$ by
\begin{equation}
L= \frac {N_i}{\sigma _i}.
\label{lum}
\end{equation}
A simultaneous measurement of the total number of events $N_t$ and the number
of elastic events $N_{el}$ is used to define the luminosity as
\begin{equation}
L= \frac {1+\rho _0^2}{16\pi }\frac {N_t^2}{dN_{el}/dt\vert _{t=0}}.
\label{lum1}
\end{equation}
The measured total cross section is independent of luminosity:
\begin{equation}
\sigma _t= \frac {16\pi }{1+\rho _0^2}\frac {dN_{el}/dt\vert _{t=0}}{N_t}.
\label{stlu}
\end{equation}

The elastic scattering amplitude must satisfy the general properties of 
analiticity, crossing symmetry, and unitarity. The unitarity of the $S$-matrix, 
$SS^+$=1, imposes certain requirements on it. In the $s$-channel, we have 
\begin{eqnarray}
{\rm Im}A(p,\theta )= I_2(p,\theta )+F(p,\theta )= \nonumber  \\
\frac {1}{32\pi ^2}\int \int d\theta _1
d\theta _2\frac {\sin \theta _1\sin \theta _2A(p,\theta _1)A^*(p,\theta _2)}
{\sqrt {[\cos \theta -\cos (\theta _1+\theta _2)] 
[\cos (\theta _1 -\theta _2) -\cos \theta ]}}+F(p,\theta ).
\label{unit}
\end{eqnarray}
The region of integration in (\ref{unit}) is given by the conditions
\begin{equation}
\vert \theta _1 -\theta _2\vert\leq \theta ,       \;\;\;\;\;
\theta \leq \theta _1 +\theta _2 \leq 2\pi -\theta .
\label{integr}
\end{equation}
The integral term represents the two-particle intermediate states of the 
incoming particles. The function $F(p,\theta )$ represents the shadowing 
contribution of the inelastic processes to the elastic scattering amplitude. 
Following \cite{hove}, we call it the overlap function. It determines 
the shape of the diffraction peak and is completely non-perturbative.
Only some phenomenological models can claim to describe it. 

In the forward direction $\theta$=0, this relation, in combination with 
optical theorem (\ref{sigt}), reduces to the  general 
statement that the total cross section is the sum of cross sections of elastic 
and inelastic processes:
\begin{equation}
\sigma _t=\sigma _{el}+\sigma _{inel}.
\label{tein}
\end{equation}

Unitarity relation (\ref{unit}) has been successfully used 
\cite{anddre, anddre1, adg, dnec} for the model-independent description of the 
Orear region between the diffraction cone and hard parton scattering, which 
became the crucial test for phenomenological models.

Experimentally, all characteristics of elastic scattering are measured as
functions of the energy $s$ and transferred momentum $t$. However, it is
appealing to have concrete information on the geometric structure of scattered
particles and the role of different space regions in the scattering process.
We should use the Fourier-Bessel transform to obtain the correspondence between
the transferred momenta and these space regions. The transverse distance between 
the centers of colliding particles, called the impact parameter $\bf b$, 
determines the effective transferred momenta $t$. The amplitudes in the
corresponding representations are related as
\begin{equation}
h(s,b)=\frac {1}{16\pi s}\int _{t_{min}=-s}^0dtA(s,t)J_0(b\sqrt {-t}).
\label{hsb}
\end{equation}
More peripheral collisions
with large $b$ lead to smaller transferred momenta $\vert t\vert $.

The amplitude $A(s,t)$ can be connected to the eikonal phase 
$\delta (s,\bf b)$ and to the opaqueness (or blackness) $\Omega (s,\bf b)$ 
at the impact parameter $\bf b$ by the Fourier-Bessel transformation
\begin{equation}
A(s,t=-q^2)=\frac {2s}{i}\int d^2be^{i{\bf qb}}(e^{2i\delta (s,\bf b)}-1)=
2is\int d^2be^{i{\bf qb}}(1-e^{-\Omega (s,\bf b)}). 
\label{eik}
\end{equation}
The integration is over the two-dimensional space of the impact parameter 
$\bf b$. 

Assuming $\Omega (s,\bf b)$ to be real and using Eq. (\ref{sigt}), we obtain
\begin{equation}
\sigma _t=4\pi \int _0^{\infty }(1-e^{-\Omega (s,\bf b)})bdb.
\label{cstb}
\end{equation}
Also,
\begin{equation}
\sigma _{el}=2\pi \int _0^{\infty}(1-e^{-\Omega (s,\bf b)})^2bdb,
\label{cselb}
\end{equation}
and
\begin{equation}
B=\frac {\int _0^{\infty }(1-e^{-\Omega (s,\bf b)})b^3db}
{2\int _0^{\infty}(1-e^{-\Omega (s,\bf b)})bdb}.
\label{Bb}
\end{equation}

To apply the inverse transformation, we must know the amplitude $A(s,t)$ at all
transferred momenta. Therefore, it is necessary to continue it analytically to 
the unphysical region of $t$ \cite{adac}. This can be done \cite{isla}.
Correspondingly, the mathematically consistent inverse formulae generally 
contain the sum of contributions from the physical and nonphysical parts of 
the amplitude $A(s,t)$. Unitarity condition (\ref{unit}) involves only the 
amplitude in the physical region; only this part of its Fourier-Bessel transform 
is also important in the unitarity relation for the impact parameter 
representation. It is written as
\begin{equation}
{\rm Im}h(s,b)= \vert h(s,b)\vert ^2+F(s,b),
\label{unib}
\end{equation}
where $h(s,b)$ and $F(s,b)$ are obtained by the direct transformation of 
$A(s,t)$ and $F(s,t)$ integrated only over the physical transferred momenta 
from $t_{min}$ to 0.
They show the dependence of the intensity of elastic and inelastic interactions 
on the mutual impact parameter of the colliding particles. Analogously to 
relation (\ref{tein}), the integrals over
all impact parameter values in this relation respectively represent 
the total, elastic, and inelastic cross sections. 
It is especially simple to calculate the overlap function from 
algebraic Eq. (\ref{unib}) if the real part is small in some subregion, i.e.
$\vert h(s,b)\vert \approx {\rm Im}h(s,b)$. Then
\begin{equation}
{\rm Im}h(s,b)\approx \frac {1}{2}(1-\sqrt {1-4F(s,b)}).
\label{imb}
\end{equation}                         
In the region where the transformed overlap function is small, $F(s,b)\ll 1$, 
the imaginary part is also small: ${\rm Im}h(s,b)\approx F(s,b)$.

However, the 
accuracy of the unitarity condition in $b$-representation (\ref{unib}) is still 
under discussion \cite{adac, isla, klok, kkl}, because some corrections due to
the unphysical region enter there, even though their role may be negligible.
Moreover, the further use of the approximate formulas of the quasi-eikonal 
unitarization often leads to failure in describing the differential 
cross section outside the diffraction cone.

The average values of the impact parameters for all -- elastic and inelastic --
processes can be estimated from the amplitude $A(s,t)$ if we assume
that $d\rho /dt=0$ at $t=0$ \cite{klok}:
\begin{equation}
<b^2(s)>_{tot}= \frac {\sigma _{el}}{\sigma _t}<b^2(s)>_{el}+
\frac {\sigma _{in}}{\sigma _t}<b^2(s)>_{in}=2B(s,0), 
\label{bav}
\end{equation}
where, e.g.,
\begin{equation}
<b^2(s)>_{el}=4\int _{t_{min}}^{0}dt\vert t\vert \left \vert 
\frac {d}{dt}A(s,t)\right \vert ^2/\int _{t_{min}}^{0}dt\vert A(s,t)\vert ^2. 
\label{abav}
\end{equation}

Nevertheless, the problem of the relative contributions of the central (small 
$b$) and peripheral (large $b$) regions under elastic hadron collisions is 
still widely disputed.
We must be especially careful when considering unitarity condition 
(\ref{unib}) with small impact parameters for certain models. Slight variations 
of $h(s,b)$ in this region may lead to strong variations of the amplitude 
$A(s,t)$ at large $\vert t\vert $.

The elastic scattering at extremely small angles allows
estimating the forward ratio of the real part of the amplitude to 
its imaginary part $\rho _0$ in experiment.
For completeness, we show an approximate expression for the 
amplitude $A(s,t)$ in the region dominated by the Coulomb amplitude and its 
interference with the nuclear amplitude:
\begin{equation}
A^{CN}(s,t)=\mp \frac {8\pi \alpha }{\vert t\vert }sf_1(\vert t\vert )
f_2(\vert t\vert )e^{i\alpha \Phi }+(i+\rho _0(s))s\sigma _te^{Bt/2},
\label{inter}
\end{equation}
where the upper (lower) sign corresponds to the scattering of particles with
the same (opposite) electric charges, the form factors of two colliding 
particles $f_j(\vert t\vert )$ added by hand in Eq. (\ref{inter}) take their 
internal composition into account, $\Phi $ is the Coulomb phase, and
$\alpha = 1/137 $ is the fine structure constant. The expressions for 
$f_j(\vert t\vert )$
and $\Phi $ depend on various prescriptions for them obtained with different
assumptions concerning the internal structure of a hadron. The most popular
shapes of the form factors are either the Gaussian fall-off with an increasing
angle, like $\exp (2t/\Lambda ^2)$, similar to that in (\ref{diff}), or the 
dipole (power-like) approximation, like $(1-t/\Lambda ^2)^{-2}$, with some more 
complicated subleading factors. The phase $\Phi $ usually contains a term with 
the typical logarithmic dependence on the angle $\theta $, which becomes large 
at very small angles, and some subleading terms. In both cases, the subleading
terms have to contain additional free parameters for a more accurate description
of experimental data. As we see, the ratio $\rho (s,t)$ in (\ref{rho}) is 
approximated by $\rho (s,0)=\rho _0$ in the fit (\ref{inter}). This implies 
that both real and imaginary parts of the nuclear amplitude exhibit the same
purely exponential $t$-dependence in the interference region (with the dominance 
of the imaginary part for small $\rho _0$). More details can be found in 
\cite{bethe, rix, sol, loch, west, cahn, sel, block, kund, deje, ppp, kope}.

\section{Where do we stand now?}

We first discuss the asymptotic properties of fundamental characteristics such 
as the total cross section $\sigma _t$, the elastic
cross section $\sigma _{el}$, the ratio of the real part to the imaginary part 
of the elastic amplitude $\rho $, and the width of the diffraction peak $B$
at infinite energies. Then we compare this with some trends in
present experimental data. 

More than half a century ago, it was claimed \cite{froi, mart} that 
according to the general principles of field theory and ideas about
hadron interactions, the total cross section cannot increase with energy 
faster than $\ln ^2s$. The upper bound was recently improved \cite{mar2},
with the coefficient in front of the logarithm shown to be half that
in the earlier limit,
\begin{equation}
\sigma _t\leq \frac {\pi }{2m_{\pi }^2}\ln^2(s/s_0),
\label{asymp}
\end{equation} 
where $m_{\pi }$ is a pion mass. 

If estimated at present energies, this bound is still much higher than the 
experimentally measured values of the cross sections, with $s_0$=1 GeV$^2$
chosen as a "natural" scale. Therefore, this is only a 
functional constraint; it forbids extremely fast growth of the total cross 
section, asymptotically exceeding the above limits. Both the coefficient
in front of the logarithm in (\ref{asymp}) and the constancy of $s_0$ are often
questioned. In particular, some possible dependence of $s_0$ on the energy $s$
has been pointed out (see, e.g., \cite{azim}).

The Heisenberg uncertainty relation shows that such a regime favors an
exponentially bounded spatial profile of the matter density distribution 
$D(r)$ in colliding particles, such as $D(r)\propto \exp (-mr)$. Because the 
energy density
is $ED(r)$ and there should be at least one created particle with mass $m$
in the overlap region, the condition $ED(r)=m$ leads to
$r\leq \frac {1}{m}\ln (s/m^2)$ and, consequently, to the functional dependence 
in (\ref{asymp}). 

It was namely Heisenberg who first proposed such a behavior
of total cross sections \cite{heis}. He considered the pion production
processes in proton-proton collisions as a shock wave problem governed by some
nonlinear field-theory equations.

To study the asymptotic regime, some theoretical arguments based on the general 
principles of field theory and the analogy of strong interactions to massive 
quantum electrodynamics \cite{chen} were promoted. The property that the limits 
as $s\rightarrow \infty $ and $M\rightarrow 0$ (where $M$ is the photon 
mass) commute has been used \cite{cw2}, implying that the asymptotic domain of 
strong interactions coincides with the massless limit of quantum electrodynamics. 
These studies led to the general geometric picture of two hadrons 
colliding with asymptotically high energies and
interacting as Lorentz-contracted black disks (see also review paper
\cite{mbrc}). In what follows, we discuss
some other possibilities as well. But as a starting point for further
reference, we describe the predictions of this proposal.

The main conclusions are:

1. For black ($\Omega (s,\bf b)\rightarrow \infty $) and logarithmically 
expanding disks with finite radii $R$ ($R=R_0\ln s,\; R_0$=const), it follows 
from (\ref{cstb}) that $\sigma _t$ asymptotically approaches infinity as
\begin{equation}
\sigma _t(s)=2\pi R^2+O(\ln s); \;\;\;\; R=R_0\ln s; \;\;\;\; R_0={\rm const}.
\label{cst}
\end{equation}

2. The elastic and inelastic processes give equal contributions 
to the total cross section:
\begin{equation}
\frac {\sigma _{el}(s)}{\sigma _t(s)}=
\frac {\sigma _{in}(s)}{\sigma _t(s)}=\frac {1}{2}\mp O(\ln ^{-1}s).
\label{frcs}
\end{equation}
This quantum mechanical result differs from "intuitive"$\;$ classical
predictions. 

3. The width of the diffraction peak $B^{-1}(s)$ must shrink because
its slope increases as (see also \cite{kino})
\begin{equation}
B(s)=\frac {R^2}{4}+O(\ln s).
\label{wid}
\end{equation}

4. The forward ratio of the real part to the imaginary part of the amplitude
$\rho _0$ must vanish asymptotically as
\begin{equation}
\rho _0=\frac {\pi }{\ln s}+O(\ln ^{-2}s).
\label{rho0}
\end{equation}
This result follows directly from Eq. (\ref{rhodi}) for 
$\sigma _t\propto \ln ^2s$.

5. The differential cross section has a shape resembling the classical 
diffraction of light on a disk:
\begin{equation}
\frac {d\sigma }{dt}=\pi R^4\left [\frac {J_1(qR)}{qR}\right ]^2,
\label{bdis}
\end{equation}
where $q^2=-t$.

6. The product of $\sigma _t$ with the value $\gamma $ of $\vert t\vert $ at
which the first dip in the differential elastic cross section occurs is a
constant independent of the energy:
\begin{equation}
\gamma \sigma _t=2\pi ^3\beta _1^2+O(\ln ^{-1}s)=35.92 {\rm \; mb}\cdot {\rm GeV}^2,
\label{gams}
\end{equation}
where $\beta _1=1.2197$ is the first zero of $J_1(\beta \pi )$.

These are merely a few conclusions among many others, albeit model-dependent ones.

None of these asymptotic predictions have been observed yet in experiment.

Surely, there is another possibility -- more realistic at present energies --
 that the black disk model 
is too extreme and the gray fringe always exists. It opens the way to  
much speculation, with many new parameters concerning particle shape and 
opacity (see, e.g., \cite{cyan, bour, ilp, desg, gbw, petr, dl84, bloc, 
kovn, trot, mesh, deje, kope}).

The black disk limit might be unrealistic. Therefore, in Table 1 we show 
the predictions of the gray disk model with the steep rigid edge described by 
the Heaviside step-function and the Gaussian disk model. The total cross 
section, the slope $B$, the ratio of the elastic to total cross section 
$X=\sigma _{el}/\sigma _t$, the ratios $Z=4\pi B/\sigma _t$ and $X/Z$, and 
the product $XZ$ are displayed there; $\Gamma (s,b)$ is the diffraction profile 
function. 

The slope $B$ is completely determined by the size of the interaction region 
$R$. Other characteristics are sensitive to the blackness of disks $\alpha $.
In particular, the ratio $X$ is proportional to $\alpha $.

The ratio $Z$ plays an important role for fits at larger angles, as explained 
in Section 4.2. It is inversely proportional to $\alpha $.
The corresponding formulas are given by (\ref{cstb}), 
(\ref{cselb}) and (\ref{Bb}). The black disk limit follows from the gray disk
model at $\alpha =1$. For a Gaussian distribution of matter, the disk 
becomes nontransparent at its center in this limit.

\medskip
\begin{table}
\medskip
Table 1. $\;\;$ The gray and Gaussian disks models $\;\; 
(X=\sigma _{el}/\sigma _t, \; Z=4\pi B/\sigma _t) $
\medskip
   
\begin{tabular}{|l|l|l|l|l|l|l|l|}
        \hline
       Model &$1-e^{-\Omega }=\Gamma (s,b)$&$\sigma _t$&
$B$&$X$&$Z$&$X/Z$&$XZ$ \\ \hline
       Gray &$\alpha \theta (R-b);0\leq\alpha<1$&$2\pi \alpha R^2$&
$ R^2/4$&$\alpha /2$&$1/2\alpha$&$\alpha ^2$&$1/4$\\ 
       Gauss &$\alpha e^{-b^2/R^2}; 0\leq \alpha \leq 1$&$2\pi \alpha R^2$
&$R^2/2$&$\alpha /4$&$1/\alpha$&$\alpha ^2/4$&$1/4$ \\
        \hline
    \end{tabular}
\end{table}
The parameter $XZ$ is constant in these models and does not depend on the
nucleon transparency. On the contrary, the parameter $X/Z$ is very sensitive 
to it, being proportional to $\alpha ^2$. Therefore, it would be extremely 
instructive to obtain knowledge about them from experimental data. 

In Table 2 we show how the above ratios evolve with energy according to 
experimental data. Most primary entries there, except the last two, are taken 
from Refs \cite{cyan, chao} with the simple recalculation $Z=1/4Y$. The data at 
Tevatron and LHC energies are taken from Refs \cite{am, toteml, totemh}. All 
results are for $pp$-scattering, except those at 546 and 1800 GeV for $p\bar p$
processes which should be close to $pp$ at these energies. The accuracy of the 
numbers listed in Table 2 can be very approximately estimated to be better 
than $\pm 10\%$ from known error bars for the cross sections and the slopes. 

\medskip
\begin{table}
\medskip
Table 2.  $\;\;$ The energy behavior of various characteristics of elastic scattering.
\medskip

    \begin{tabular}{|l|l|l|l|l|l|l|l|l|l|l|l}
        \hline
$\sqrt s$, GeV&2.70&4.11&4.74&6.27&7.62&13.8&62.5&546&1800&7000\\ \hline
   X&0.42&0.28&0.27&0.24&0.22&0.18&0.17&0.21&0.23&0.25 \\  %\hline
   Z&0.64&1.02&1.09&1.26&1.34&1.45&1.50&1.20&1.08&1.00 \\  %\hline
   X/Z&0.66&0.27&0.25&0.21&0.17&0.16&0.11&0.18&0.21&0.25 \\ %\hline
   XZ &0.27&0.28&0.29&0.30&0.30&0.26&0.25&0.26&0.25&0.25\\  \hline
   
\end{tabular}
\end{table}
The most interesting feature of the experimental results is the minimum of the 
blackness parameter $\alpha $ at ISR energies.
It can be clearly seen in the minima of $X$ and $X/Z$ and in the maximum of $Z$
at $\sqrt s$=62.5 GeV. The steady decrease of ratios $X$ proportional to 
$\alpha $ and $X/Z$ proportional to $\alpha ^2$ up to the ISR energies and  
their increase at S$p\bar p$S, Tevatron, and LHC energies means that the 
nucleons become more transparent up to the ISR energies and more black toward 
7 TeV. The same conclusion follows from the behavior of $Z$, which is inversely
proportional to $\alpha $. The value of $Z$ rapidly approaches its limit for the 
Gaussian distribution of matter in the disk. For the Gaussian shape, the 
parameter $X/Z$ cannot exceed 0.25. This model is excluded only at low 
energies. According to Eq. (\ref{stseb}) $XZ\approx 0.25(1+\rho _0^2)$, which is 
indeed close to 0.25 within the limits of experimental errors, the estimate of 
$\rho _0^2\leq 0.02$, and slight variations of $B$ inside the cone in the
framework of our crude model as predicted in Table 1. This shows that our models 
are not bad for qualitative estimates in a first approximation. 

Before dwelling on various fits, we briefly comment on some  
important general trends in high-energy data observed in experiment. 

1. Total cross sections increase with energy. At present energies, the 
power-like approximation is the most preferable one. The preasymptotic behavior 
of $\sigma _t$ proposed in earlier papers \cite{chen, cw2} was
\begin{equation}
\sigma _t \propto s^a\ln ^{-2}s,
\label{pre}
\end{equation}
where the numerical value of $a$ was estimated to be of the order of unity in 
strong interactions. It was shown in \cite{cww} to lie in the range between
0.08 and 0.2, which is close to values obtained in recent phenomenological fits.
The power growth persists in a wide interval of energies (see Ref. \cite{famesi} 
for the recent analysis of experimental data). Consequently,
the density distribution in colliding particles is closer to a power-like
dependence than to an exponential one in that energy range.

2. The ratio $\sigma _{el}/\sigma _t $ decreases from low energies to those of 
ISR, where it becomes approximately 0.17 and then strongly increases 
up to 0.25 at LHC energies. However, it is still quite far from the 
asymptotic value 0.5, corresponding to the black disk limit. 

The only higher-energy data came from the Pierre Auger collaboration, 
which recently reported \cite{auger} a measurement
of the inelastic $p$-air cross section $\sigma _{in}^{p-air}$ at
$\sqrt s=57\pm 6$ TeV. After some corrections and Glauber model calculations, it
results in the $pp$ inelastic cross section $\sigma _{in}^{pp}\approx 90$ mb. 
Some models \cite{mbfh05, blha} extrapolate their predictions for the total 
cross section to this energy and obtain a value of about 135 mb. Hence, the 
ratio of the inelastic to the total cross section could become equal to 0.67, 
which is smaller than 0.75 at 7 TeV. However, it is premature to reach any 
definite conclusions because of large errors in the cosmic ray data and the 
underestimated value of the total cross section predicted at 7 TeV by the
model \cite{mbfh05, blha}. The extrapolation to infinite energies done in
the same model leads to this ratio estimated as 0.509, which is compatible with 
the black-disk predictions. Still, asymptopia is but an elusive concept!

Sometimes, the modified black disk limit is attributed to the sum of elastic and
diffractive processes \cite{gpps}. It may then be that
\begin{equation}
\frac {\sigma _{el}+\sigma _{diff}}{\sigma _t}\rightarrow \frac {1}{2},
\label{gpps}
\end{equation}
where $\sigma _{diff}$ is the sum of cross sections of single and 
double inelastic diffraction. The fits in Ref. \cite{famesi} suggest separately
the relations
\begin{equation}
\frac {\sigma _{el}}{\sigma _t}\rightarrow \frac {1}{3} \;\; \rm{and} \;\;
\frac {\sigma _{diff}}{\sigma _t}\rightarrow \frac {1}{6}.
\label{gpps1}
\end{equation}

3. The diffraction peak shrinks about twice from energies
about $\sqrt s \approx 6$ GeV, where $B \approx 10$ GeV$^{-2}$, to the LHC
energy, where $B \approx 20$ GeV$^{-2}$. At ISR energies, the slope $B(s)$
increases logarithmically. Accounting for LHC data requires a stronger 
dependence than a simple logarithmic one. The terms proportional
to $\ln ^2s$ are usually added in phenomenological fits. Even then, 
predictions \cite {okor, caok} are not completely satisfactory. 
At present energies, in connection with the power-like preasymptotic behavior 
of $\sigma _t$, we could also expect a faster-than-logarithmic shrinkage of 
the diffraction peak. 

The tendency in peak behavior at larger $\vert t\vert $ also changes
with an energy increase. In the energy region up to ISR, it becomes less
steep near its end (see Figs 4, 5 in Ref. \cite{zrt}), but its slope
increases at the LHC energies. Both the minimum and maximum following the peak 
shift to smaller $\vert t\vert $. 

As regardss the behavior of the differential cross section in the function of the
transverse momentum behind the maximum, the $t$-exponential of the diffraction 
peak is replaced, according to experimental data, by the 
$-\sqrt {\vert t\vert }\approx -p_t$-exponential at the intermediate angles:
\begin{equation}
d\sigma /dt\propto e^{-2a\sqrt {\vert t\vert}}, \;\;\; a\approx \sqrt B.
\label{orear}
\end{equation} 
The slope $2a$ in this region also increases with energy, and the whole Orear
region shifts to the ever lower transferred momenta. 

In this connection, we also note an intriguing property of the ratio 
$Z=4\pi B/\sigma _t$, which is closely related to the value of the slope $B$.
From Table 2, we see that it is about 1 at $\sqrt s$ = 4 GeV, increases to 
1.5 at ISR energies and then again drops to around 1 at 7 TeV. This ratio, in 
combination with values of $\rho $ at different angles, determines the slope in 
the $\vert t\vert $-region beyond the diffraction peak at any $s$ (see
Ref. \cite{anddre1} and the discussion in the subsection 4.2.2). According to 
Eqs (\ref{cst}), (\ref{wid}), $Z$ should decrease and be asymptotically equal 
to 1/2 for the black disk limit, such that the relation
\begin{equation}
\sigma _t=8\pi B
\label{sas}
\end{equation}
be asymptotically fulfilled. At the LHC energy 7 TeV, the coefficient in
the right-hand side is still half as much. However, if the preasymptotic 
power-like increase in the total cross section accompanied by a slower increase 
in the slope persists, the tendency to this limit looks quite promising. 

The relation between $\sigma _t$ and $B$ is also discussed in Refs 
\cite{famesi, fmen}. In particular, the fits in \cite{famesi} correspond
to the value $Z\approx 0.93$ at Auger energies 57$\pm $6 TeV, i.e. lower 
than 1 at 7 TeV.

4. As a function of energy, the ratio $\rho _0$ increases from negative values
at comparatively low energies, crosses zero in the region of hundreds GeV
and becomes positive at higher energies. This is a general tendency for 
collisions of any initial particles. For $pp$ scattering, the prediction of 
(\ref{rho0}) with values of $s$ scaled by 1 GeV  is still 
somewhat higher (about 0.177) than the estimates from dispersion relations 
($\approx $0.14 in Refs \cite{blha, drna}), even at 7 TeV, 
while strongly overshooting them at ISR, where $\pi /\ln s \approx 0.37$.
No logarithmic decrease is seen in these predictions, which, however, depend on 
the behavior of the total cross section at higher energies. Moreover, the value 0.14
can only be reached according to (\ref{rho0}) at the energy of 75 TeV. Probably,
at energies higher than 75 TeV, the first signs of approach to the asymptotic 
regime will become visible. No data about $\rho _0$ at the LHC energies exist 
yet. The local value of $\rho _0$ estimated from Eq. (\ref{rhodi}) with a 
power-like fit of the total cross section, proportional to $s^{\Delta }$, is 
$\rho _0\approx \pi \Delta /2$. That agrees quite well with the soft Pomeron
intercept $\Delta \approx 0.08$.

5. To describe the shape of the differential cross section in the diffraction 
cone, significant corrections to Eq. (\ref{bdis}) must be added at present 
energies. This is discussed in subsection 4.1.

6. The product $\gamma \sigma _t$ changes from 39.5 mb$\cdot $GeV$^2$ at 
$\sqrt s=6.2$ GeV to 51.9 mb$\cdot $GeV$^2$ at $\sqrt s=7$ TeV
and strongly deviates from the predicted asymptotic value (\ref{gams}).
The total cross section $\sigma _t$ increases faster than $\gamma $ decreases.

From the geometrical point of view, the general picture is one of protons 
becoming blacker, edgier and larger (BEL) \cite{rhpv}. 
We conclude that even though the qualitative trends may be considered rather 
satisfactory, we are still quite far from the asymptotic regime, even at  
LHC energies. This feature may be connected \cite{dren} with the strong evolution 
of the parton content of strong interactions at present energies, revealing 
itself in an increase in the number of active parton pairs inside each proton 
with energy increase (higher density) and a softening of the structure functions,
which leads to lower energy shares $x$ for each parton pair (larger radii).

\section{Experimental data and phenomenological models}

As always, our knowledge about particular physical processes is limited by the
practical possibility of measuring their characteristics. As mentioned above,
numerous experimental data on the elastic scattering of hadrons at various angles 
and at different energies have been obtained. Unfortunately, in some of them 
the available region of angles is strongly limited by the experimental setup.
Therefore, a comparison with theoretical proposals is possible only in the
corresponding range of angles and energies.

The data and their fits at various energies and in different intervals of 
transferred momenta for different participating particles
are so numerous that it is impossible to show all of 
them in a single review paper. Therefore, from the very beginning, we use 
the latest results of the TOTEM collaboration at the highest 
LHC energy, 7 TeV, as a reference point \cite{toteml, totemh}. The discussion 
of theoretical models is also concentrated near these data.

The total and elastic cross sections at 7 TeV are respectively estimated as 
98.3 mb and 24.8 mb.\footnote{Here, we do not 
reproduce the statistical and systematic errors. They are shown in the 
original papers.}. The figures from published papers \cite{toteml, totemh} 
demonstrating the behavior of the differential cross section as function 
of the transferred momentum are displayed below. 
They clearly confirm the existence of the three regions discussed above.

The cross section shape in the region of the diffraction cone \cite{toteml} is 
shown in Fig. 1. The $ t $-exponential behavior with $B\approx$20.1 GeV$^{-2}$ 
is clearly seen at $\vert t\vert <0.3$ GeV$^2$. The peak steepens at the end of 
the diffraction cone, and its slope becomes approximately equal to 
23.6 GeV$^{-2}$ in the $\vert t\vert $ interval of (0.36 -- 0.47) GeV$^2$. 
The results at somewhat larger angles \cite{totemh} in the Orear region are 
presented in Fig. 2. The dip at
$\vert t\vert \approx 0.53$ GeV$^2$ with a subsequent maximum at 
$\vert t\vert \approx 0.7$ GeV$^2$ and the $\sqrt {\vert t\vert }$-exponential 
behavior are demonstrated. Some curves, corresponding to different model 
predictions, are also drawn here. The same data as in Fig. 2 are 
shown in Fig. 3, but with more details, including the steepend slope, the dip 
position, and the region of $\vert t\vert ^{-8}$-behavior. The last one 
is ascribed to the hard parton scattering processes. 

We congratulate all members of the TOTEM collaboration with this fantastic 
achievement! Their efforts are truly appreciated when estimating the values of 
angles at which the measurements had to be done. They were even smaller than 
10$^{-4}$! Detectors had to be installed at very long distances from the 
collision point to obtain results at low transferred momenta. These data 
revived interest to elastic scattering.  

Theoretical models usually describe the diffraction cone and 
values of total and elastic cross sections related to it more or less 
precisely (therefore, their fits, that are almost indistinguishable in 
that region, are not drawn in Fig. 1). 
However, all of them fail to quantitatively predict the behavior of the 
differential cross section outside the diffraction cone as can be seen in 
Fig. 2. The predictions of 
five models \cite{blha, bour, isla2, jll, ppp} are drawn here. They are very 
widely spread around the experimental line. We can conclude that
just this region becomes the Occam razor for all models. In what follows, we 
consider these models, as well as some others, in more detail. 

\begin{figure}
\includegraphics[width=\textwidth, height=8cm]{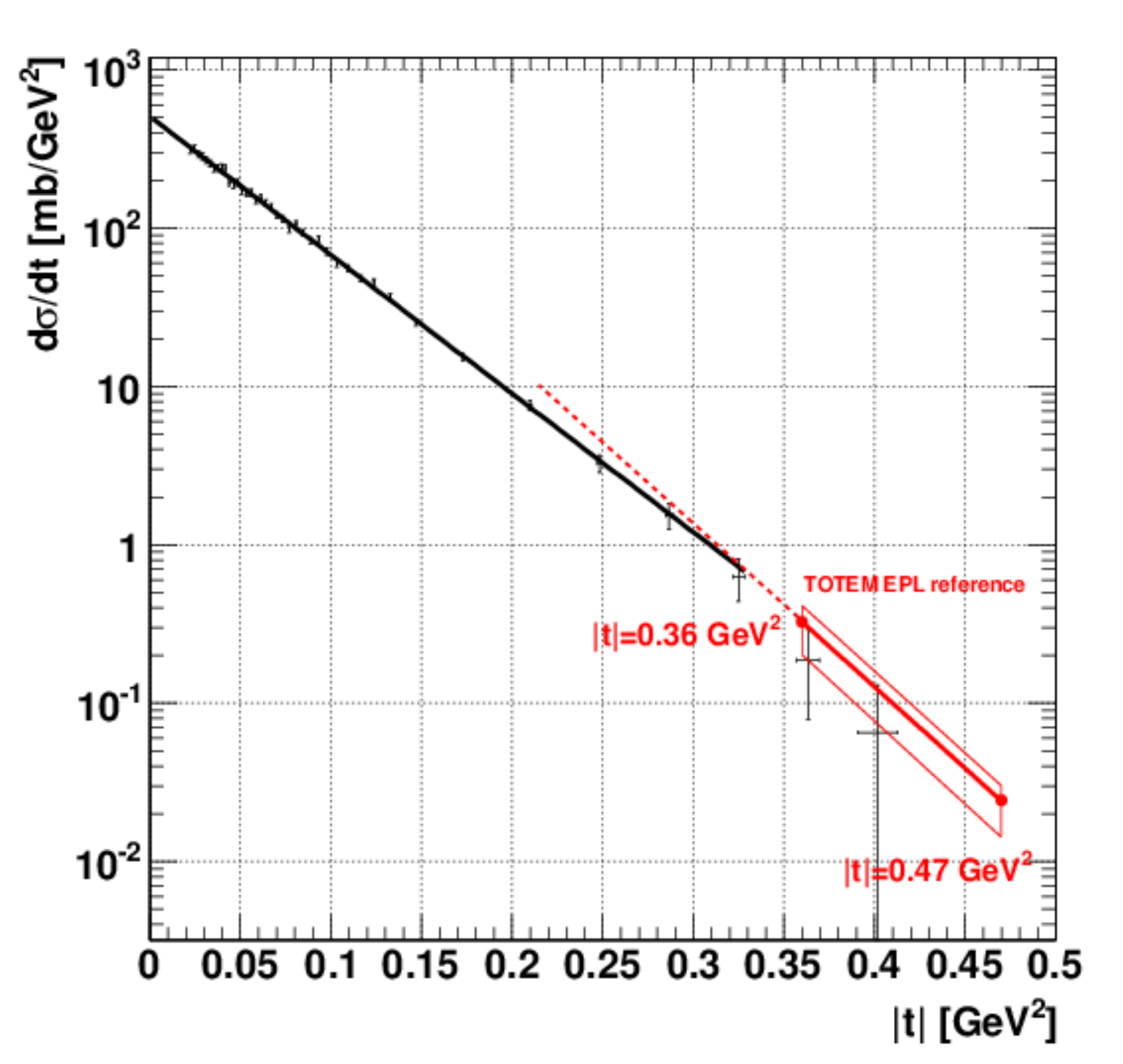}

Fig. 1. The differential cross section of elastic proton-proton scattering at 
$\sqrt s$=7 TeV measured by the TOTEM collaboration  
(Fig. 4 in \cite{totemh}). \\
The region of the diffraction cone with the $\vert t\vert $-exponential 
decrease is shown.

\end{figure}

\begin{figure}
\centerline{\includegraphics[width=\textwidth]{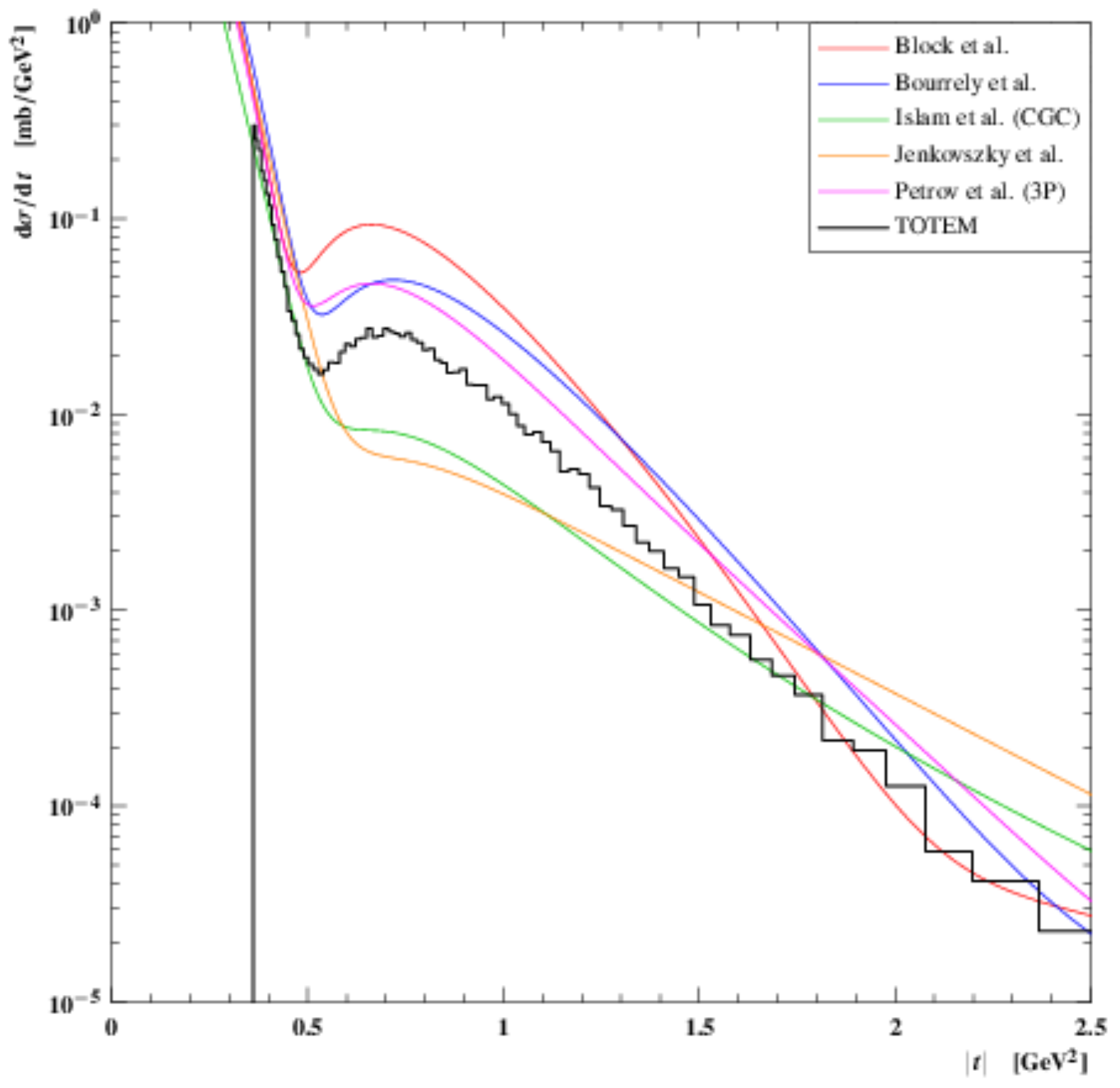}}

Fig. 2. The differential cross section of elastic proton-proton scattering at 
$\sqrt s$=7 TeV measured by the TOTEM collaboration  
(Fig. 4 in \cite{toteml}). \\
The region beyond the diffraction peak is shown. The predictions of five models
are demonstrated. 

\end{figure}

\begin{figure}
\centerline{\includegraphics[width=\textwidth]{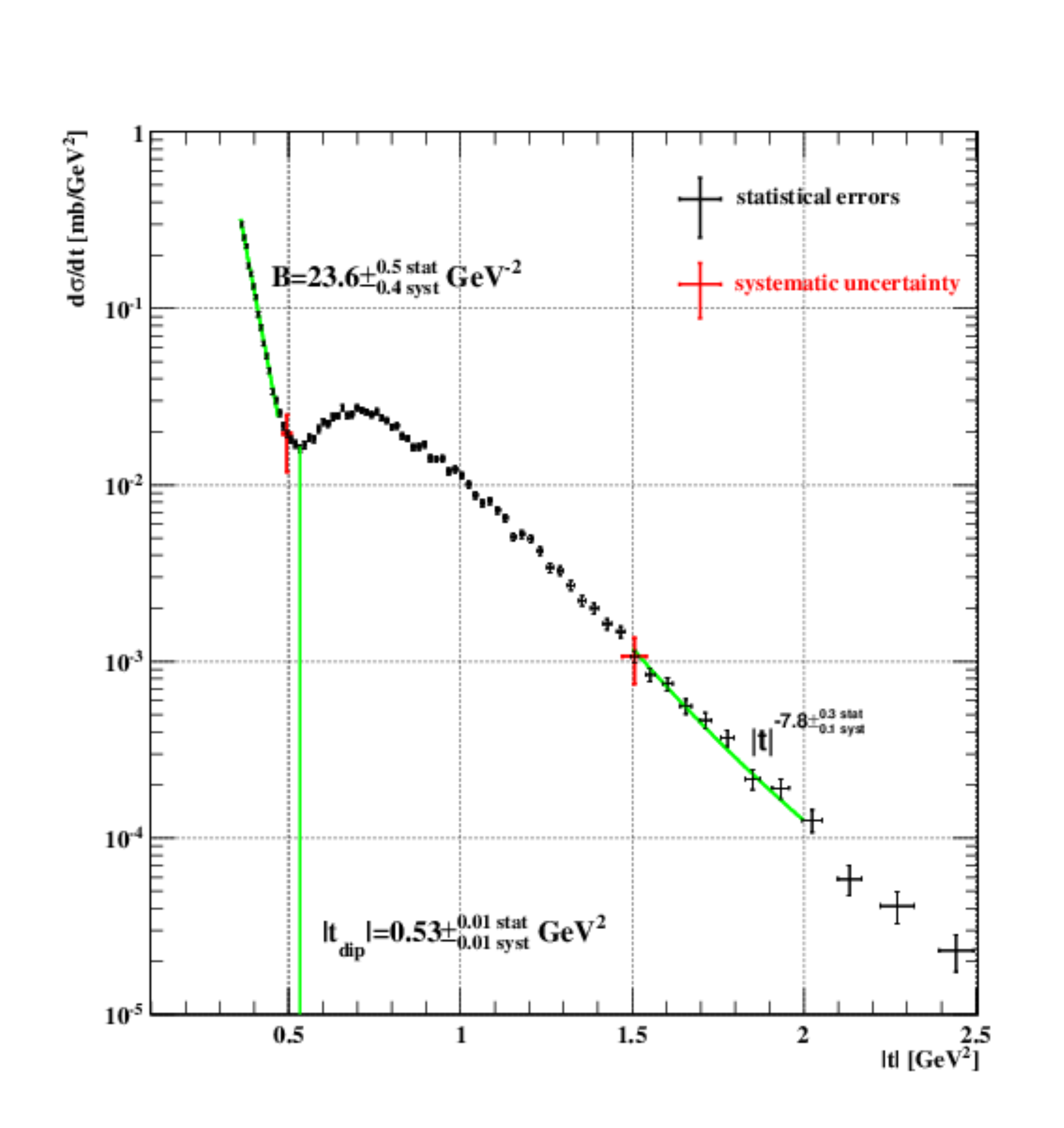}}

Fig. 3. The differential cross section of elastic proton-proton scattering at 
$\sqrt s$=7 TeV measured by the TOTEM collaboration 
(Fig. 3 in \cite{toteml}). \\
The same regions as in Fig. 2 are shown with the values of the steepened slope
at the outskirts of the diffraction peak, the position of the dip and the
power-like behavior at the largest transferred momenta.

\end{figure}

The three intervals of $\vert t\vert $ (the diffraction cone, the Orear regime,
and the region of hard parton scattering) are characterized by different 
dynamical content, as we understand it now. They require separate 
approaches to their descriptions.
It seems reasonable that these regions are regulated by different but
interrelated physical mechanisms. In particular, different spatial regions
of overlapping colliding objects are responsible for corresponding effects.
Three subsections 4.1, 4.2 and 4.3 are devoted to theoretical approaches
to their explanation.                                        

\subsection{Diffraction cone and geometrical approach}

The internal structure of colliding, strongly interacting particles plays a 
crucial role in the outcome of their collisions. In high-energy hadron-hadron
scattering, each hadron behaves as an extended object. They can be described 
by their size and the density of their constituents. The simplest models are
demonstrated in Table 1.

Since long ago, it has been believed that hadrons contain some denser core 
surrounded by a meson (pion) 
cloud at their periphery. This idea was a cornerstone of the one-pion exchange 
model, which was first proposed in Ref. \cite{dche} to describe particle 
production in peripheral interactions. It evolved into the well-known 
multiperipheral and (multi)Reggeon exchange models ( see, e.g., 
\cite{amati, akim, ddun} for early review papers). They are rather successful 
in describing many features of multiparticle production processes. The 
multiperipheral approach developed, for instance, in the framework of the 
Bethe-Salpeter equation (see Ref. \cite{ddun}) can be considered an attempt 
to account for the $t$-channel unitarity. 

Nowadays, it is commonly believed that, at very high 
energies, the total cross section is dominated by peripheral events. 
In modern parlance, this is related to the long-range 
nature of the field of "perturbatively massless" gluons. The exchanged boson 
mass may mimic a nonperturbative mass gap in QCD with the "effective" gluon mass 
of the order of 1 GeV and a gluon-gluon correlation length about 0.3 fm. The 
pion mass scale is rather small, and more general "boson" exchange is 
preferred. The weight factors of different mass scales take  
the impact parameter distribution of the particle opacity into account.

The role of inelastic channels in describing elastic scattering can be 
revealed by understanding the origin and prescribing a definite shape to the 
overlap function $F(p, \theta )$ in the $s$-channel unitarity condition 
(\ref{unit}) or, equivalently, to its Fourier transform in the impact parameter 
picture. The scattering is mainly diffractive, i.e., it is due to the 
absorption of incoming waves in many open inelastic channels. Its 
quantitative field-theory treatment presents a serious unsolved problem. 

The overlap function
contains the sum of products of a matrix element of the inelastic process
with a particular final state and the complex conjugate matrix element with
the same final particles content. However, their kinematical difference must be 
taken into account, due to the fact that the two final protons are scattered 
at an angle $\theta $ relative to the initial ones. Correspondingly, the 
overlap of the momentum distributions of the intermediate inelastic $n$-particle
states is nontrivial kinematically and, what is especially important,
the phases of these matrix elements become crucial. The phases are related to 
the position in space where particles are produced. It has been pointed out
in many papers \cite{fuku, zale, mich, giff} that only the phase cancellation 
effect, which is closely related to particle correlations in inelastic processes,
can lead to a realistic shape of the diffraction cone. The problem of 
properly accounting for them has not yet been solved. 

At the same time, elastic scattering should be less peripheral because of 
a larger number of exchanged objects if regarded as an $s$-channel iteration
of the overlap function. 
The great difficulty in transferring large momenta reveals itself already in the 
sharp shape of the forward diffraction peak. There have been numerous attempts 
to understand it in terms of the peripheral approach (see, e.g., 
\cite{sopk, dar, dur, arno, jack}). Unfortunately, no framework for commonly
accounting both the $s$- and $t$-channel unitarity conditions has been 
developed.

In general, there have been many ideas proposed for describing elastic scattering 
processes, but no cogent theoretical arguments to justify the particular 
forms relying mainly on "intuition" have been offered. The fact that they are 
very simple is usually the only advantage. Any strict interpretation is an 
idealization and as such it should not be expected to be exactly true. 

\subsubsection{Geometry of the internal hadron structure}

The key elements of the geometric approach are: the use 1) of the impact 
parameter picture with Fourier-Bessel transformation (\ref{eik}),
(\ref{hsb}) from the transferred momenta amplitude to the spatial description,
2) of eikonal approximation (\ref{eik}), and 3) of unitarity condition 
(\ref{unib}). 
The $S$-matrix in the impact parameter picture is chosen in the exponential form
\begin{equation}
S(s,b)=e^{-\Omega (s,b)}
\label{so}
\end{equation}
and the convolution approximation for the real opacity $\Omega $ for elastic
$AB$ scattering is used:
\begin{equation}
\Omega (s,b)=KD_A\otimes D_B.
\label{osb}
\end{equation}
Here, $\otimes $ denotes the convolution of hadronic matter density 
distributions $D$ for $A$ and $B$. $K$ is an energy-dependent factor.
The assumptions about the validity of the eikonal approximation, the nearly 
imaginary character of the scattering amplitudes at low transferred momenta, 
the proportionality between the hadronic matter distribution and the electric 
charge distribution, the exponentiation of the $S$-matrix in $b$-space, and the 
validity of unitarity condition (\ref{unib}) are widely used.

The droplet model \cite{wuy, byy} for elastic collisions was the first to 
fully exploit all the above elements. Particles were pictured as very much similar 
to nuclei. Correspondingly, the notion of the density distribution $D$ inside a
particle was introduced such that
\begin{equation}
\Omega (s,b)={\rm const}\int _{-\infty }^{+\infty }D((b^2+x^2)^{1/2})dx.
\label{dens}
\end{equation}
In potential models it corresponds to the WKB approximation. For the Gaussian 
shape of $h(s,b)$, it is possible to solve for $D$ from (\ref{dens}), obtaining 
the function familiar in the theory of Bose-Einstein condensation of free 
particles \cite{byy}. In the droplet model, the properties of the disk are 
independent of the energy at sufficiently high energies. Many diffractive
minima in the differential cross section have been predicted. The dipole form 
factors in the $t$-representation led to $\Omega (s,b)$ with a shape 
of the modified Bessel functions, which 
allowed fitting differential cross sections at ISR energies \cite{chou}.
The intuitive picture of high-energy hadron collisions as two extended objects
breaking in fragments (and thus defining the overlap function!) has promoted
the hypothesis of limiting fragmentation \cite{bcyy} inspired by the droplet 
model.

Models based on consideration of tower diagrams \cite{chen, hctt}
predict that the disk becomes larger and more absorptive as energy increases.
Both the black core and gray fringe expand with energy and become more 
absorptive.

The first estimates of the radii of protons, pions, and kaons from their form 
factors \cite{ttc, ttch, ttcy} showed that protons are larger than pions and 
kaons. This is not surprising in view of the smaller cross sections of $\pi p$ and
$Kp$ interactions than those of $pp$. The typical size is somewhat
smaller than 1 fm. The proton hadronic matter distribution was fitted by a
dipole form similar to the electric form factor but with the energy-dependent
radius. 

Other early attempts to consider elastic scattering of hadrons also stemmed from 
the analogous simple geometrical treatment of their internal structure 
\cite{kris, chou, isl, cwu}. Later, more complicated models were used. The main 
focus is, surely, on processes at small angles within the
diffraction cone. They define the bulk contribution to the elastic scattering
cross section due to the steep fall-off of the distribution with increasing
angles. Different models happen to fit experimental data in the cone quite well 
in a wide energy range. But they fail outside the diffraction peak, as mentioned
above. Large-angle scattering requires more central collisions with a lower
impact parameter to probe the internal content of particles.
Therefore, these regions of transferred momenta are discussed separately 
below. 

Some ideas stemmed from regularities in inelastic processes. The multiplicity
distributions of created particle
are closely related to the purely geometric notion of the centrality 
of collisions. When the scaling of multiplicity distributions \cite{kno} was
supported by experimental data, the proposal of the geometric scaling 
\cite{ddd} for the elastic amplitude was promoted. The difficulties in
accelerating the various parts of a nucleon without breaking it up had to be
accounted for.

The basic idea of the geometric scaling is that, at sufficiently
high energies, the amplitude $A(s,t)$ depends on a single variable, the
scaling parameter $\tau $:
\begin{equation}
\tau = -\frac {t}{t_0}\ln ^2\frac {s}{s_0}.
\label{ddd}
\end{equation}
This idea has led to several predictions at asymptotically high energies and is
still actively being debated now. Such scaling was proved \cite{akm71, budi74} 
for cross sections increasing as $\ln ^2s/s_0$ and for an infinitesimally small 
ratio of the real to imaginary part of the amplitude $\rho \rightarrow 0$
at $s\rightarrow \infty $. The latest results on $\rho (s,t)$ discussed in 
subsection 4.2.5 do not support this assumption.

The purely geometrical standpoint is adopted in Refs \cite{isla, ilp, isla2}.
The three regions in the behavior of the differential cross section are clearly
reflected in the three spatial scales of the internal hadronic structure
considered in \cite{ilp,isla2}. The authors of this three-scale model claim 
that the nucleon has an outer cloud of the quark-antiquark condensate, an inner 
shell of the baryonic charge density, and a still smaller internal core of massless 
color-singlet valence "quarks" surrounded by low-$x$ gluon clouds  
about 0.3 fm in size. This picture is shown in Fig. 4.

\begin{figure}
\includegraphics[height=8cm]{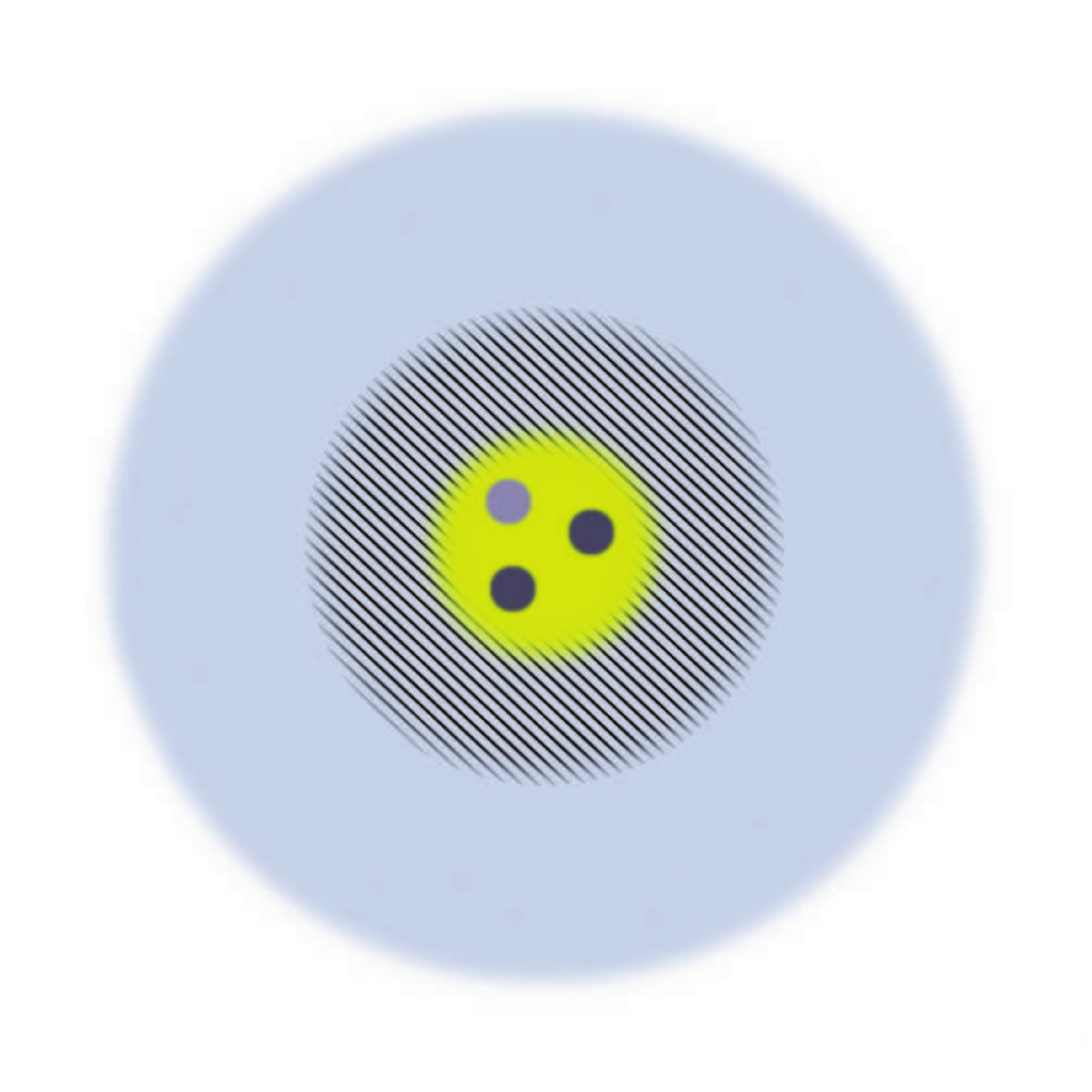}

Fig. 4. The nucleon structure according to the model \cite{isla, ilp, isla2}.\\
The three regions of the internal structure are supposed to be directly 
responsible for the three regimes in the behavior of the differential cross 
section.

\end{figure}

The diffraction 
cone is described as a result of cloud-cloud interaction, represented by a
class of potentials containing the sum of the modified Bessel functions. The 
least massive exchanged quanta are the most important ones. At larger momentum
transfers, the baryonic charge at intermediate distances is probed by the
$\omega $-exchange. The internal region filled in by the valence quarks starts 
playing its role in the presence of even larger transferred momenta. 

The diffraction profile function, which defines the range of different
densities and, correspondingly, different forces, is taken to be
\begin{equation}
\Gamma (s,b)= 1-\Omega (s,b)= g(s)\left [\frac {1}{1+e^{(b-r)/a}}+
\frac {1}{1+e^{(-b+r)/a}}-1\right ].
\label{gisl}
\end{equation}
The parameters $r$ and $a$ are energy dependent,
\begin{equation}
r=r_0+r_1(\ln (s/s_0)-i\pi /2); \;\;\;\; a=a_0+a_1(\ln (s/s_0)-i\pi /2),
\label{risl}
\end{equation}
and $g(s)$ is a coupling strength; $s_0$=1 GeV$^2$. 

These functions render the
shape of the differential cross section, similar to the Fraunhofer diffraction
(see subsection 4.1.2), with the form factor proportional to 
$\pi dq/\sinh \pi dq$ ($q^2=-t,\; d$ is an adjustable parameter)
proposed a long time ago \cite{frv, frah, spma}. This form factor also extends 
somewhat to transferred momenta outside the diffraction cone.
Unfortunately, the contemporary phenomenological analysis of experimental data
is not able to determine the impact parameter profiles unambigously.

The scattering due to $\omega $-exchange is parameterized by the product of the 
$\omega $-propagator and two form factors $F$ directly in the 
$(s,t)$-representation:
\begin{equation}
A_{\omega }(s,t)\propto se^{i\chi (s,0)}\frac {F^2(t)}{m_{\omega }^2-t}.
\label{omisl}
\end{equation}

The amplitude due to quark-quark scattering has two "structure factors" $G$ of 
valence quarks (different from the above form factors!), the propagator with the
black disc radius $r_B$ of $qq$ asymptotic scattering and $s$-dependent factors 
with the hard Pomeron intercept equal to $1+\alpha _h$:
\begin{equation}
A_{qq}(s,t)\propto ise^{i\chi (s,0)}(se^{-i\pi/2})^{\alpha _h}
\frac {G^2(t)}{r_B^{-2}+\vert t\vert }.
\label{qqisl}
\end{equation}
In total, there are seventeen adjustable parameters in the model.

As mentioned in \cite{totemh}, the fits according to this model predict too low 
value of the slope $B$ at $\vert t\vert $=0.4 GeV$^2$ and strongly disagree
with experiment at 7 TeV outside the diffraction peak (see Fig. 2). 
Formulas (\ref{omisl}), (\ref{qqisl}) are aimed to improve the fit just in this
region, but they do not help.

In general, an internal region of the nucleon where the gluons cluster
around the original valence quarks resembles the valon model \cite{hwa, hwaz}.
Similar pictures arise in the QCD-inspired models discussed below.

Surely, some care should be taken for any such model to be accepted and the
geometric picture to be considered seriously, especially in view of its success or
failure to describe experimental data in the whole range of transferred 
momenta at various energies.

\subsubsection{The modified Fraunhofer diffraction}

For a long time (see, e.g., Ref. \cite{sing}), the formulas of classical 
diffraction of light on a (black or grey) disk with the traditional Bessel 
functions have been used for hadronic reactions. Recently, an analogous 
expression for the elastic amplitude was considered in Ref. \cite{ugal}:
\begin{equation}
A(s,\vert t\vert=q^2)=C\frac {dq}{\sinh (\pi dq)}\left [i\frac {J_1(Rq)}{Rq}+
\frac {\rho }{2}J_0(Rq)\right ].
\label{ugal}
\end{equation}
The free parameters in arbitrarily chosen analytic expression (\ref{ugal})
are $C,\; R,\; d,\; \rho $. The first term resembles the
expression for the black disk (\ref{bdis}). The suppression at large 
transferred momenta is assumed to be approximated by the form factor in front
of the Bessel functions. 
In the impact parameter representation this shape corresponds to the ordinary 
Fermi profile used, e.g., in Refs \cite{bddd, ilp, isla2} and shown in 
Eq. (\ref{gisl}):
\begin{equation}
h(b)\propto \frac {1}{1+e^{(b-R)/d}}.
\label{ferm}
\end{equation}
The second term in (\ref{ugal}) in brackets takes the contribution due to the 
real part of the amplitude into account. It should smooth the behavior of the 
differential cross section near zeros of the first term. This seems to be the 
only difference from the first component of the previously discussed model
\cite{ilp, isla2}.

And, again, comparison with experimental data shows that the results of fits 
are satisfactory in the diffraction cone, but not outside it. The form factor
in front of common Bessel functions does not fit the large $\vert t\vert $
trends of experimental distributions.

Throughout these developments, modifications of early guesses have been found
necessary, but the general spirit of the geometrical description remains 
immutable and viable.

\subsubsection{Electromagnetic analogies}

The strongly interacting content of hadrons is often considered to be 
similar to their electromagnetic substructure \cite{chou}. Similarly to the
droplet model, the assumption of the proportionality between the hadronic 
matter distribution and the electric charge distribution is used in many
models. However, in most of them, the electromagnetic form factors are used 
in combination with Reggeon exchanges because, considered alone, they do not 
reproduce the energy dependence of the main characteristics. However, the 
assumption about the full congruence of these distributions is not necessarily 
valid, since gluons do not carry an electric charge even though they play an 
important (if not decisive) role in high-energy strong interactions. That is 
why the charge and matter distributions in some models are parameterized 
separately or some corrections are added.

Using the experience from calculation of tower diagrams in
electrodynamics and the impact-parameter representation, it was proposed 
\cite{bour, bour1, bour2, bour3} that the possibility of choosing 
the opacity $\Omega (s,{\bf b})$ in a factorized form be considered:
\begin{equation}
\Omega (s,{\bf b})= R(s)F({\bf b}^2)+({\rm non-leading \;\; terms}),
\label{fact}
\end{equation}
where $R(s)$ is chosen to be crossing symmetric under $s\leftrightarrow u$ and 
to reproduce the energy dependence of the Pomeron, considered as a fixed Regge 
cut,
\begin{equation}
R(s)=\frac {s^c}{(\ln s)^{c'}}+\frac {u^c}{(\ln u)^{c'}},
\label{ss}
\end{equation}
while $F({\bf b}^2)$ is taken as the Bessel transform of
\begin{equation}
F(t)=f\vert G(t)\vert ^2\frac {a^2+t}{a^2-t}.
\label{ftfb}
\end{equation}
Here, $G(t)$ stands for the proton "nuclear form factor", parameterized like the
electromagnetic form factor with two poles:
\begin{equation}
 G(t)=\frac {1}{(1-t/m_1^2)(1-t/m_2^2)}.
\label{gel}
\end{equation}
Other factors with the parameter $a$ are introduced "by hand". They can be 
treated just as a correction due to the different shapes of distributions of
charge and matter. There are six 
adjustable parameters in total used at high energies if the Regge background
is neglected. The noticeable $t$-dependence of the slope $B(t)$ in the 
diffraction cone is predicted. However, its values at 7 TeV are lower than 
experimental ones (about 18 GeV$^{-2}$ instead of 20.1 GeV$^{-2}$) at 
$10^{-2}<\vert t\vert <0.3$ GeV$^2$, 
slightly exceed them in the tiny interval near 0.35 GeV$^2$ and do not
reach the value 23.6 GeV$^{-2}$ mentioned above.

This model is close to the TOTEM data \cite{toteml, totemh} for the dip 
position and the exponential at very large $\vert t\vert $, but predicts  
values of the differential cross section in the Orear range, 
$\vert t\vert \geq$ 0.36 GeV$^2$, about twice as large (see Fig. 2). 
In addition to the dip,
some "oscillations"$\;$ at the transferred momenta of several GeV$^2$ are
predicted (up to the energy 6000 TeV) but not yet observed. In general,
such structures appear as a byproduct of the eikonal approach and 
unitarization procedure (see, e.g., Ref. \cite{desg}). Their energy dependence
is strongly determined by the parameters used in formula (\ref{ss}) to 
account for the crossing symmetry property of the amplitude.

The same parameters are crucial for the behavior of the real part of the
amplitude. It is interesting
that the model predicts the dominance of the imaginary part of the amplitude
even at large transferred momenta. The real part becomes important only at
zeros of the imaginary part. The dip and oscillations are noticeable precisely 
there. Near the cone, the model predicts two zeros of the real part of the full 
(Coulomb + nuclear) amplitude at $\vert t\vert =$0.0064 GeV$^2$ and the nuclear 
amplitude alone at $\vert t\vert \geq$ 0.18 GeV$^2$, as well as one zero of 
the nuclear imaginary part at $\vert t\vert =$0.5 GeV$^2$. In the differential 
cross section, the last zero is partly filled in by the real part.

We note the difference between the power-like expression for $F(t)$ and its
exponential behavior in the traditional Regge models. The exponentiation
of this form of $F(t)$ leads to additional oscillations.

The similar but more complicated combination of the form factors has been used 
in Refs \cite{sel, cuse, sete, sel12}. The authors consider the $t$-dependent 
Mellin 
transforms of parton distributions and claim that the first moment defines the
form factor of the standard Pomeron $G$, while the second moment $H$ corresponds
to interaction attributed to three nonperturbative gluons. Thus, the behavior
of the differential cross section at small $t$ is determined by the
elactromagnetic form factors and by matter distribution at large $t$.
The Born term of the elastic scattering amplitude is written as
\begin{equation}
A^{Born}(s,t)=h_1 G^2(t)F_a(s,t)(1+r_1/{\hat s}^{0.5})+
h_2H^2(t)F_b(s,t)(1+r_2/{\hat s}^{0.5}),
\label{ase}
\end{equation}
where
\begin{equation}
F_a(s,t)={\hat s}^{\epsilon _1}e^{B(s)t}; \;\;\;\;
F_b(s,t)={\hat s}^{\epsilon _1}e^{B(s)t/4},
\label{fab}
\end{equation}
\begin{equation}
G(t)=\frac {L_1^4}{(L_1^2-t)^2}\frac {4m_p^2-2.793t}{4m_p^2-t}; \;\;\;\;
H(t)=\frac {L_2^4}{(L_2^2-t)^2}.
\label{gh}
\end{equation}
$L_1^2=0.71$ GeV$^2$, $\; L_2^2=2$ GeV$^2$, $\; {\hat s}=se^{-i\pi /2}/s_0, \;
s_0=1$ GeV$^2$, $\; B(s)=\alpha '\ln s/s_0, \; \alpha '=0.24$ GeV$^{-2}$.
We note that the slope of the second term is chosen as one fourth of 
the first term. The final form of the amplitude is obtained after 
eikonalization of the Born contribution using the opacity
\begin{equation}
\Omega (s,b)=\frac {1}{2\pi}\int d^2qe^{i \bf q \bf b}A^{Born}(s,q^2=-t).
\label{fafb}
\end{equation}
The total cross section at 7 TeV was predicted to be equal to 95 mb.
Authors demonstrate good fits of $pp$ and $p\bar p$ differential cross 
sections, as well as of $\rho _0(s)$, in a wide energy range, including the
TOTEM data. Only five (three for high energies and two for low energies)
adjustable parameters are claimed to be used if all above values are regarded 
as fixed. In fact, there are 10 such additional "hidden" parameters in total if 
the hard Pomeron
is also considered. Surely, the contribution from secondary Reggeons at LHC 
energies is negligible, i.e., smaller than the experimental errors.

The real part of the hadron amplitude is completely determined by the complex 
expression for $\hat s$. Its $t$-dependence appears just as a byproduct of the
eikonalization procedure. As a function of $t$, it tends to zero
at $\vert t\vert \approx $0.16 GeV$^2$ at the energy of 7 TeV. 
The interesting predictions of the $t$-behavior of $\rho (s,t)$ at
nonforward transferred momenta for different energies are presented. 
They are discussed in more details in subsection 4.2.4.

\subsubsection{Reggeon exchanges}

The Regge-pole model is beyond dispute one of the most explored.
It has already been noticed that the notion of Regge trajectories has been used
in the preceding subsections as well. The only reason to discuss these models 
there separately was their stronger inclination to the use of nonexponential 
electromagnetic form factors and geometric pictures in the $s$-channel
approach. At the same time, Reggeon models appeal mostly to the $t$-channel
approach.

The amplidutes with Reggeon exchanges in the $t$-channel are the natural 
candidates for explaining
the exponential decrease of the differential cross section (\ref{diff}) with 
the squared transferred momentum $\vert t\vert$ inside the diffraction cone. 
Just this shape is typical for them, because it follows from the linearity
of Regge trajectories. Moreover, they predict the logarithmical 
increase of the hadronic radii as the energy increases, i.e., the logarithmical 
increase in the cone slope $B$ or the corresponding shrinkage of the width of 
the diffraction cone. This prediction is also supported by experiment.
In the common Regge-pole models, the disk becomes larger and slightly more 
transparent as energy increases.

The standard Regge-type models \cite{jenk, petr, sel, koko} use the combination 
of contributions due to the exchange by the (multicomponent) Pomeron, Odderon, 
and secondary Reggeon trajectories corresponding to $f$ and $\omega $ mesons 
with or without the form factors chosen in a simple exponential form or as
power-like expressions resembling the electromagnetic structure of colliding 
partners discussed in the preceding subsection. The price to be paid is the 
increased number of adjustable 
parameters at each step of sophistication. To be more or less realistic, one
has to use the knowledge about some of them from other (independent?) 
experimental results. But even under this condition, the ambiguity of their
choice and sensitivity to fitted parameters leave some freedom in the conclusions.

The amplitudes of $pp$ and $p\bar p$ scattering are approximated by the sum
of terms corresponding to the leading (Pomeron and Odderon) and nonleading 
($f$ and $\omega $ meson) Regge trajectories:
\begin{equation}
A(s,t)^{pp}_{p\bar p}=A_P(s,t)+A_f(s,t)\mp [A_{\omega }+A_O(s,t)],
\label{appbar}
\end{equation}
where the labels $P,\;f,\;O,\;\omega $ stand for the relevant contributions. The 
sign in the $pp$ and $p\bar p$ amplitudes differs for $C$-even and $C$-odd terms.

The contributions of the nonleading Regge poles are written as
\begin{equation}
A_R(s,t)=a_Re^{-i\pi \alpha _R(t)/2}e^{b_Rt}(s/s_0)^{\alpha _R(t))}
\label{areg}
\end{equation}
with $\alpha _R(t)=a_R+b_Rt$.

While the secondary trajectories are usually chosen in a standard linear way, 
the Pomeron and Odderon contributions can be regarded, for example, 
as dipoles with nonlinear trajectories \cite{flpr, ghpr, angr, jenk, fjop}
\begin{equation}
A_P(s,t)=i\frac {a_Ps}{b_Ps_0}[r_1^2(s)e^{r_1^2(s)(\alpha _P-1)}-\epsilon _P
r_2^2(s)e^{r_2^2(s)(\alpha _P-1)}],
\label{apom}
\end{equation}
where $r_1^2(s)=b_P+L-i\pi /2, \;  r_2^2(s)=L-i\pi /2, \; L=\ln(s/s_0).$
The unknown Odderon contribution is assumed to be of the same form as that of
the Pomeron. The parameters of the trajectories and of the absorption 
$\epsilon _P$ need to be adjusted. Their nonlinearity may be connected with the
two-pion threshold following from the $t$-channel unitarity \cite{flpr, angr,
kmrys}.
However, there could be double counting of the graphs with Pomerons attached
on both sides to the pion loop. This is well known from old peripheral models
of inelastic processes, where the self-consistent Bethe-Salpeter equation had 
to be used for the proper account of the pion-nucleon vertices. Different forms 
of nonlinear trajectories are in use. For instance,
the Pomeron trajectory is chosen in \cite{bert} with four free parameters as
\begin{equation}
\alpha (t)=\alpha _0-\gamma \ln (1+\beta \sqrt {t_0 -t}).
\label{bert}
\end{equation}
The more complicated nonlinearity was used in \cite{kmrys}. However,
the use of the pion mass as a scale there is questionable in view of the 
above discussion.

The origin of the Pomeron and the parameterization of its trajectory are still
being debated. There is no strict rule for choosing its shape. The dipole and even 
tripole forms of unitarized Pomeron have been attempted. They mimic cut 
contributions \cite{luni, jema, cuma, marty}. 

Moreover, there are arguments in favor of two Pomerons with different 
intercepts. Even the fits with three Pomerons are sometimes used 
\cite{ppp, petr}. The soft Pomeron contributes a term with the energy
dependence $s^{a_s}, \;\; a_s\approx 0.08$ to hadron-hadron total cross 
sections, and the hard Pomeron makes a small
(at present energies) contribution with a stronger energy dependence 
$s^{a_h}, \;\; a_h\approx 0.4$. These values of the intercepts stem from the 
discussions of HERA data (see, e.g., \cite{dla}).  
Although the hard-Pomeron exchange was unnecessary for describing  
hadron-hadron total cross sections up to energies $\sqrt s$ below 1 TeV, it
may reveal itself at LHC energies, as argued in \cite{dola}. The model in
\cite{dola} uses only two terms in the expansion for opacity:
\begin{equation}
\Omega (s,b)=\Omega _S(s,b)-\frac {\lambda }{2}\Omega _S^2(s,b),
\label{dola}
\end{equation}
where $\Omega _S$ stands for the contribution from single exchanges of Reggeons
(two Pomerons, $f$, and $\omega $) with the adjustable parameter $\lambda $ as well 
as for the triple-gluon exchange of the form $Cst^{-4}$ needed at larger values
of $ \vert t\vert $ and matched at some $t=t_0$ to exponential shapes of the
diffraction peak and to the dip region. Certainly, adding such a term allows
fitting the total cross section value at 7 TeV, but there is a suspicion
that the sharp increase of the hard-Pomeron contribution will  overpredict 
the cross sections at higher energies. The unitarization will become 
mandatory once again. The quality of the fit of the differential cross section 
beyond the diffraction peak is no better than of those fits shown in Fig. 2.

Several variant forms of Born amplitudes and different kinds of eikonalization
have been attempted. There is no consensus on their choice. 

The form of the eikonal similar to (\ref{dola}) is chosen in \cite{golo} with
the exponential suppression
\begin{equation}
\Omega _S(s,b)=A_0\exp [-m(s)(r_0^2+b^2)^{1/2}]
\label{golo}
\end{equation}
for central interactions. The peripheral part of the Pomeron interaction with
the meson cloud is parameterized \cite{golo} by a small term increasing with the
energy and resulting in a $\sqrt {\vert t \vert }$ exponential fall-off of
the differential cross section. The geometric picture corresponds to a
black disk with a grey fringe, similarly to the above-described model 
\cite{bour}.

In general, it is not easy to estimate the total number of the adjustable 
parameters in different models. 
There are parameters related either to $t$-
($b$-) or $s$-dependence. In some papers, it is often assumed that part 
of them are known from fits of other characteristics of hadron or 
electromagnetic interactions at various energies and can
therefore be considered known beforehand.
 
For example, it is claimed that the model in \cite{jenk} contains about 15 
parameters. In this case, it is quite difficult to find the proper minima for 
the matrix of $\chi ^2$ values. It is well known how unstable the final 
results can be: one has to choose the step-by-step procedure for doing this and 
use some special computer codes. 

There are 25 adjustable parameters shown in Table 1 in papers 
\cite{petr, ppp}. However, they include some values assumed to be a'priori fixed
in \cite{jenk}. At the same time, additional form factors were inserted 
in the formulas, albeit with preliminary "fixed" parameters.
They were used to fit 982 $pp$ and $p\bar p$ data points
in a wide energy range. Besides the elastic differential cross sections, the 
total cross sections and the ratios $\rho _0$ were considered. The fit in the 
interference region of Coulomb and hadronic amplitudes with the same parameters
helped in choosing among the different Coulomb phases proposed previously.

A similar situation is seen in the fits in Ref. \cite{sel}, where it is 
claimed that the number of parameters is much less (5 only!). However, there are
many others (in particular, concerning the energy behavior and form factors) 
that are hidden parameters. They are held fixed from the very beginning, as was 
discussed in subsection 4.1.3. 

As mentioned before, the simple exponential form of the differential cross 
section in the diffraction cone is quite well described. This becomes possible 
mainly due to the $t$-shape of the Pomeron trajectory (\ref{apom}) and other 
Reggeons contributions in (\ref{areg}). The fits in this region at the different 
energies shown, e.g., in \cite{jenk, petr, sel}, 
are quite impressive. The evolution of the diffraction 
cone slope with energy is reproduced (as described by $L$ in (\ref{apom})). 
Unfortunately, the variety of forms of Pomeron trajectories with different 
intercepts, slopes, and shapes of residues unitarized in different ways and/or
substituted by Regge-cuts is so large that it is impossible to show all of 
them in this review due to the limited space. 

The cuts with nonlinear 
trajectories mimic hard scattering \cite{dl84}. A 
common problem appears in predicting them at larger angles. The fit
according to the model in \cite{petr, ppp} seems to be most successful in 
predicting the position of the dip and the shape at large $\vert t\vert $
but exceeds the absolute value approximately twofold. The model in \cite{jenk}
strongly underestimates it, with the wrong position of the dip and much slower
decrease at $\vert t\vert >1.5$ GeV$^2$. This is well 
demonstrated in Fig. 2 and is also discussed below.                                               

We mention that all these papers follow the general approach proposed
much earlier \cite{angr, donn}. They just deal with more detailed fits of newly
available experimental data.

\subsubsection{QCD-inspired models}

Each incident particle consists of a superposition of Fock states with $n$ 
partons \cite{feyn}, which are scattered instantaneously and simultaneously 
by the other particle. Some QCD-inspired models using this statement have 
been developed. The role of partons is played by quarks and gluons.

The two competing mechanisms of hadron interactions, the increase in the density
($\alpha $ in Table 1) and in the radius $R$, determine their specific
features. In QCD, they can be respectively ascribed to the leading order 
solution of the BFKL equation \cite{bfkl} and to the long range 
(Weizs{\"a}cker-Williams) nature of the field of massless gluons. The density 
increase due to the BFKL-evolution leads to a power-like increase of the total 
cross section, which is nonunitary and violates Froissart bound (\ref{asymp}). 
Therefore, at the critical density of the order $1/\alpha _S$, 
the density saturation must be taken into account
\cite{glr}. The QCD evolution in all orders in the gluon density but in the
leading logarithmic approximations is treated by the JIMWLK equations 
\cite{jimwlk}. With account of multiple scattering effects, they can be
simplified in the large $N_c$ limit to a single nonlinear BK equation for the
gluon density \cite{bako} when the induced field density is small. 

The density growth effects are preasymptotic. According to \cite{kovn}, they are 
described by a hard Pomeron, while the growth of the size of the black 
saturated regions (the radius) is attributed to a soft Pomeron. The hard 
Pomeron manifests itself
in small systems or in small subregions inside hadrons. The soft Pomeron
appears in hadronic systems of the typical size and is related to an increased 
size in the impact parameter space. Only the increase due to the perturbative 
expansion in the transverse plane remains effective. 

There is no common 
consensus about this scenario proposed in Ref. \cite{kovn}. The soft Pomeron
is often used \cite{kope, khar, kopopo} in attempts to explain the preasymptotic
power growth of cross sections by an additional nonperturbative mechanism
superimposed on the BFKL scenario of hard Pomeron. It is ascribed mainly
to the density growth of gluon clouds around quarks and not to the spatial 
scale of the interaction. Even though the size of gluon clouds increases,
it is still limited by a short separation from their source. The proton looks 
like three valence quarks surrounded by gluon clouds or spots with mean sizes 
about 0.3 fm, smaller than the proton radius, of the order of 1 fm. 
Radiation of any additional gluon from the cloud
adds the factor $\ln s/s_0$ to the interaction cross section, and hence their
sum gives a power-like term of the form
\begin{equation}
\sigma _t=\sigma _0+\sigma _{\Delta}(s/s_0)^{\Delta };  \;\;\;\; 
\Delta=4\alpha _S/3\pi \approx 0.17
\label{kppp}
\end{equation}
with a large constant term $\sigma _0$ and small $\sigma _{\Delta }$. Using the
standard dipole form factors of protons and quasieikonal unitarization in the
impact parameter space, the authors of this two-scale model \cite{kope, kopopo} 
fit many distributions 
with 10 parameters for $t$-dependence (subject to 2 additional constraints) and 
some parameters for the $s$-dependence. Such fits are, of course, aimed at 
high energies
of colliding protons where the effects of secondary Regge-trajectories die out.
They are mainly successful in the diffraction cone and, consequently, in 
describing the energy dependence of the total and elastic cross sections.

Such a form of the total cross section with an energy-independent term 
$\sigma _0$ was proposed a long time ago \cite{chlu, elki, gaha} and actively
developed later \cite{almv, dupi} in the framework of the parton model and
semihard QCD, with the gluon-gluon interaction playing the main role.

The main role of gluons is also incorporated in \cite{bloc, blha}. The profile 
is chosen in a form containing the $gg, \; qq, \; qg$-terms:
\begin{equation}
\Omega =\sigma _{gg}W(b;\mu _{gg})+\Sigma _{gg}(C+C_R\frac {m_0}{s^{1/2}})
W(b;\mu _{qq})
+\Sigma _{gg}C_{qg}\ln \frac {s}{s_0}W(b;(\mu _{qq}\mu _{gg})^{1/2}),
\label{eikb}
\end{equation}
where the impact parameter distribution functions are
\begin{equation}
W(b;\mu )=\mu ^2 (\mu b)^3K_3(\mu b)/96\pi , 
\label{wbm}
\end{equation}
and the gluon-gluon interaction cross section is
\begin{equation}
\sigma _{gg}=C_{gg}\int \Sigma _{gg}\Theta (\tau s-m_0^2)F_{gg}(x_1,x_2)d\tau
\label{sgg}
\end{equation}
with $\Sigma _{gg}=9\pi \alpha _S^2/m_0^2; \;\; F_{gg}=\int f_g(x_1)f_g(x_2)
\delta (\tau -x_1x_2)dx_1dx_2; \;\; f_g(x)=N_g(1-x)^5/x^{1+\epsilon }$. 

The Froissart bound for the total cross section is reproduced with
\begin{equation}
\sigma _t=2\pi (\epsilon /\mu _{gg})^2\ln ^2(s/s_0).
\label{frs}
\end{equation}
The parameter $\mu _{gg}$ describes the area occupied by gluons in the colliding 
protons (the size effect), and $\epsilon $ is defined via their gluonic structure 
functions and, therefore, controls their soft gluon content (the density 
effect).

Again, being successful in the diffraction peak with its shape and
normalization, the model in \cite{bloc, blha} fails to predict the correct behavior 
of the differential spectrum outside it \cite{toteml, totemh}.
Its prediction is more than three times larger than the experimental value at 
the dip and subsequent maximum, while falling too steeply at ever higher 
$\vert t\vert$ above 1.5 GeV$^2$ (see Fig. 2).

Attempts to consider the semihard scattering of quarks and gluons 
can be found in Refs \cite{dupi, marg, bghp, kasp}.

The traditional partonic description of the process is considered in a series
of papers \cite{kmrys, martd, rmkh, khoz}. The partonic approach with a hard 
BFKL Pomeron
can be merged into the domain regulated by the soft Pomeron. The transition from 
hard to soft is induced by absorptive multi-Pomeron effects in a limited
energy range. The evolution produces parton cascades, not strongly ordered in
transverse momenta, with hot spots of a relatively small size in $b$-space.
The saturation is driven by the enhanced multi-Pomeron graphs, also regulating 
the high-mass dissociation. The calculations are done with a 3-channel 
quasi-eikonal unitarization using the opacity formalism. They reproduce the 
shapes of the differential cross sections from ISR to LHC within the
diffraction cone.

Another picture was considered in the framework of the functional integral 
approach in Refs \cite{bena, berg, dosi, dfk} using the model of the stochastic 
vacuum and making the assumption that the proton has a quark-diquark structure of
the color dipole i.e. two quarks out of three are close together in the transverse 
directions. A matrix cumulant expansion is used for vacuum expectation values of
Wegner-Wilson loops \cite{bena} related to hadronic amplitudes. The QCD vacuum
parameters (the gluon condensate or the string tension, the vacuum correlation 
length, and the parameter due to the non-Abelian tensor structure), as well as 
the hadron size, have been used. The imaginary part of the amplitude in the
$b$-representation was calculated. Its contribution to experimentally measured
quantities was shown to describe the ISR and Tevatron data in the diffraction 
peak reasonably well. 
   
A more phenomenological approach to the quark-diquark model was attempted 
in Refs \cite{bibz}. As above, the correlated 
quark and diquark constituents are considered. According to the detailed 
analysis performed in \cite{necs} from ISR to LHC energies in the
region $0.36<\vert t\vert <2.5$ GeV$^2$, the model is able to describe the data
quite well, even outside the diffraction peak, except the narrow strip around 
the dip. But it shows a much stronger dip (by several orders of magnitude) 
there than the experimentally observed one. Moreover, similarly to the 
abovementioned calculations, the model ignores the contributions to the real 
part of the elastic scattering amplitude. As we saw previously, such 
contributions can smooth this dip. If so, their shape should drastically
differ from that of the imaginary part, at least in this strip, as happened,
for example, in the models with electromagnetic form factors \cite{bour, sel12}.

\subsection{Intermediate angles: the dip and the Orear regime}

As long ago as the 1960s, experiments on elastic $pp$- and $\pi p$-scattering at
comparatively low energies between 6.8 and 19.2 GeV in the laboratory system 
\cite{cocc, orear, alla} showed that the steep exponential fall-off of the 
differential cross section as a function of the squared transferred momentum 
$\vert t\vert $ is replaced by a slower dependence at larger $\vert t\vert $.
They showed that just after the diffraction cone a shoulder was observed
and, even more surprising, a behavior exponentially decreasing with the angle 
or with $\sqrt {\vert t\vert }$, which was called the Orear regime after  its 
investigator \cite{orea, orear}. The special session was devoted to these findings 
at the 1968 Rochester conference in Vienna. The shoulder evolved later into the
minimum or dip at higher ISR energies. It has also been observed at the LHC, 
as seen in Figs 2 and 3.

It is interesting that at FNAL-ISR energies, $\sqrt s$=6 - 60 GeV, the exponential
fall-off with an increase of $\sqrt {\vert t\vert }\approx p_t$ was observed up to
quite large values of $\vert t\vert \approx $ 10 GeV$^2$ \cite{hart, fais, zrt},
with the exponent in the range from 6.2 to 7 GeV$^{-1}$  (see Table 7 in Ref. 
\cite{zrt}). It is even larger at the LHC (about 8 - 9 GeV$^{-1}$). The region
becomes more narrow and shifts to lower values of $\vert t\vert $ from 0.5 to
1.5 GeV$^2$. The power-like regime already
shows up at about $\vert t\vert \approx $2 - 2.5 GeV$^2$ (see Fig. 3).

\subsubsection{Gaussian fits}

From the very beginning, it was noticed \cite{orea} that it is possible to fit
the differential cross sections at intermediate values of the momentum transfer
by the dependence exponential in $\sqrt {\vert t\vert }$ (or $\theta $) except
the relatively small shoulder region. To take that into account as well, it was 
primarely
proposed \cite{krisc} to use fits with Gaussian functions with alternating signs 
of the coefficients directly in the expression for the amplitude. The similar 
approach was later advocated in \cite{rhpv, hv74, h76, hv79, hmv74}.
In this way,
both the diffraction peak and larger $\vert t\vert $-behavior could be 
described. No reference to any phenomenological model is given. From the 
geometrical point of view, one can imagine an internal structure with
envelopes of alternating density.

Such an empirical approach has been recently used \cite{cm97, am08, fms11}
for fits of experimental data at ISR energies. The following parameterization
of the amplitude is proposed in Ref. \cite{fms11}:
\begin{equation}
A(s,t)=s\left [(\rho \sigma _t-\sum _{i=2}^m4\pi a_i)e^{b_1t}+
\sum _{i=2}^m4\pi a_ie^{b_it}+i(\sigma _t-\sum _{j=2}^n4\pi c_j)e^{d_1t}+
i\sum _{j=2}^n4\pi c_je^{d_jt} \right ],
\label{fms11}
\end{equation}
where $m<n$. The fits at different energies give information about the ratio of
the real to imaginary part of the amplitude $\rho (t)$, besides the values of 
adjustable parameters $a_i, b_i, c_j, d_j$. Two different methods were used.
In total, there are 14 to 16 free parameters. The results of nonlinear fits are 
rather unstable, and the conclusions are somewhat controversial. In particular, 
the numbers of zeros in ${\rm Im}A(s,t)$ and ${\rm Re}A(s,t)$ differ in these 
methods. The dominance of the real part of the amplitude at intermediate
values of the momentum transfer in one of the methods is not confirmed when
the other method is used. 

A similar fit was recently attempted and applied to TOTEM data in Refs 
\cite{tros, gpps}. The earlier proposal in Ref. \cite{baph} with 
phenomenologically chosen two $t$-exponentials and the relative interference phase 
responsible for the dip was applied to TOTEM data. Using
five parameters, it is possible to describe these data in the whole interval
of transferred momenta. We note that, similarly to the model in \cite{sel12}, the 
slope of the second exponential term is chosen several times smaller than that 
of the main term. Moreover, when the electromagnetic form factors were 
tried in place of simple exponentials, the fit became worse. 

Two exponentials without the interference term inside the diffraction peak and
a Tsallis-type distribution outside it were used in \cite{fms2}.
It was possible, with the help of nine free parameters, to fit the data at
energies from 19.4 GeV to 7 TeV.

In some way, this fit business with no reference to any theoretical model
looks more like art than science, especially if no conclusions
about the hadron structure are obtained. Such an approach will hardly be 
conclusive in the future.

\subsubsection{Phenomenological models}

Theoretical indications of the possibility of a new regime with an increase
in transferred momenta were obtained 
even earlier \cite{hove, amati, cott}. It was treated as a consequence of
the simple iteration of processes approximated by a Gaussian within the 
diffraction cone. The term $I_2$ in unitarity condition with Gaussians
inserted into the integrand gives rise to a Gaussian with a width, that is 
twice as big, i.e. to a shape twice as wide as the diffraction cone. Further
iterations lead to further widening. Therefore, multiple exchanges were 
considered. However, the 
results did not fit new experimental findings. This failure was explained as 
resulting from the improper treatment of the unitarity requirements and
incorrect choice of the overlap function. 

The droplet model relations between form factors and the elastic
amplitude for hadronic scattering at infinite energy (see Eqs (1) and (2) in
Ref. \cite{cy68}) predict a series of kinks (or zeros) in the differential
 cross section,
which could be related to dips. Dip position movement to lower $\vert t\vert $
with a growth of the total cross section was predicted in Ref. \cite{cy81}.
There is also an indication of several dips (or shoulders) at larger 
$\vert t\vert $ in the models \cite{bour, dupi, sel12} using the electromagnetic 
form factors with subsequent eikonalization (cf. Figs 2 and 11). 

In accordance with the experimental data shown in Fig. 2, only one dip is 
predicted by others. For example, 
it was described in Ref. \cite{ols} on the basis of a modified optical model
\cite{cy81}. In the framework of the geometric scaling approach \cite{dk78} 
the numerical integration of the relation
\begin{equation}
\frac {d\sigma }{dt}(s,t)/\frac {d\sigma }{dt}(s,0)=[\phi ^2(\tau )+\rho _0^2
(d(\tau \phi (\tau )/d\tau )^2]/(1+\rho _0^2),
\label{dk78}
\end{equation}
where
\begin{equation}
{\rm Im}A(s,t)=s\sigma _t\phi (\tau )  \;\;\;\;\;  \phi (0)=1
\label{imsi}
\end{equation}
was performed with $\tau $ defined by Eq. (\ref{ddd}).

It was predicted that the dip should even disappear at energies higher than 
$\sqrt s\approx 300$ GeV but, probably, can reappear again at ever higher 
energies. As we know now, it is clearly seen at 7 TeV.
The imaginary part has been chosen in such a way that it has
a zero at the dip. The absence of additional dips is
explained as the deviation of the eikonal from a simple Gaussian with some
flattening at small impact parameters (see subsection 4.2.3). That shows
strong sensitivity to the choice of tiny details of the phenomenological 
eikonal and also agrees with the properties of the 
overlap function to be discussed in more detail below. These results were 
confirmed and extended to $p\bar p$ collisions in Ref. \cite{fish}. 

Processes described by diagrams with multiple exchange by Pomerons are claimed 
to be responsible 
for the Orear regime at intermediate angles according to Ref. \cite{adya}.
The differential cross section is predicted to have the form
\begin{equation}
\frac {d\sigma }{Cdt}=\exp [-2\sqrt {2\pi \alpha '(0)\vert t\vert \xi \cot 
(\phi /2)}\varphi _1(\xi )]\cdot [1+\lambda \cos(2\sqrt {2\pi \alpha '(0)
\vert t\vert \xi \tan(\phi /2)}+\varphi _0)],   
\nonumber
\label{adya}
\end{equation}
where $\xi =\ln (s/4m^2)$, and $C,\; \phi ,\; \varphi _0,\; \varphi _1,\; 
\lambda ,\; \alpha '(0)$ are
adjustable parameters. There are oscillations directly imposed on the
exponential fall-off with the same exponent. They should be well pronounced.
So far, no such oscillations have been observed.

A less strong statement about some saturation of the diffraction cone due to 
multiple Pomeron exchanges is made in Refs \cite{kmr, glm}.

\subsubsection{Unitarity condition}

A theoretical explanation based on the consequences 
of the unitarity condition in the $s$-channel has been proposed in Refs
\cite{anddre, anddre1}. The careful fit to experimental data showed good 
quantitative agreement with experiment \cite{adg}. Nowadays, the same approach
helps explain the TOTEM findings \cite{dnec} (see Fig. 5 below).

We consider the left-hand side and the integral term $I_2$ in unitarity 
condition (\ref{unit}) at the angles $\theta $ outside the diffraction peak.
Because of the sharp fall-off of the amplitude with the angle, the 
leading contribution to the integral arises from a narrow region around the
line $\theta _1 +\theta _2 \approx \theta $. Therefore, one of the amplitudes
should be inserted at small angles within the cone as a Gaussian, while the 
other is kept at angles outside it. Integrating over one of the angles yields 
the linear integral equation:        
\begin{equation}
{\rm Im}A(p,\theta )=\frac {p\sigma _t}{4\pi \sqrt {2\pi B}}\int _{-\infty }
^{+\infty }d\theta _1 e^{-Bp^2(\theta -\theta _1)^2/2} f_{\rho }
{\rm Im}A(p,\theta _1)+F(p,\theta ),
\label{linear}
\end{equation}
where $f_{\rho }=1+\rho _0\rho (\theta _1) $.

It can be solved analytically (see \cite{anddre, anddre1} for more details)
with two assumptions: that the role of the overlap function $F(p,\theta )$ is 
negligible outside the diffraction cone and the function $f_{\rho }$ can be
approximated by a constant, i.e., $\rho (\theta _1)=\rho _l$=const. Both
assumptions are discussed in the next subsections.

It is esay to check that the eigensolution of this equation is
\begin{equation}
{\rm Im} A(p,\theta )=C_0\exp \left (-\sqrt 
{2B\ln \frac {Z}{f_{\rho }}}p\theta \right )+\sum _{n=1}^{\infty }C_n
e^{-({\rm Re }b_n)p\theta } \cos (\vert {\rm Im }b_n\vert p\theta-\phi _n)
\label{solut}
\end{equation}
with
\begin{equation}
b_n\approx \sqrt {2\pi B\vert n\vert}(1+i{\rm sign }n) \;\;\;\;\;\;\; n=\pm 1, \pm 2, ...
\label{bn}
\end{equation}
This expression contains the term exponentially decreasing with $\theta $ (or 
$\sqrt {\vert t \vert }$) (Orear regime!) with oscillations strongly damped by 
their own exponential factors imposed on it. These oscillating
terms are responsible for the dip. Just this formula was used in 
Refs \cite{adg, dnec} for fits of experimental data in a wide energy range.
The ratio $\rho $ was approximated by its average values in and outside the
diffraction cone, with $f_{\rho }=1+\rho _0\rho _l$, where $\rho _l$ is 
treated as the average value of $\rho $ in the Orear region. The fits at 
comparatively low energies \cite{adg} are consistent with $f_{\rho }\approx 1$,
i.e., with small values of $\rho _l$ close to zero. The great surprise
of the fit in \cite{dnec} of TOTEM data shown in Fig. 5 was the necessity of 
using the negative value of $\rho _l\approx -2.1$ large in modulus. 

Being model-independent, this solution suffers from some limitations that are
inherent for the unitarity relation, in general, and for the unitarity
equation (\ref{linear}), in particular. First, it predicts the 
dependence on transferred momenta $p\theta \approx \sqrt {\vert t\vert }$ but 
not the dependence on the collision energy. Second, it is applicable in a 
restricted (and not rigorously defined) range of angles in the Orear region.

The elastic scattering differential cross section outside 
the diffraction cone (in the Orear regime region) is
\begin{eqnarray}
\frac {d\sigma }{p_1dt}&= &\left (   e^{-\sqrt 
{2B\vert t\vert \ln \frac {4\pi B}{\sigma _tf_{\rho }}}}
\right.
\nonumber \\
&+&\left.
 p_2e^{-\sqrt {2\pi B\vert t\vert}} \cos (\sqrt {2\pi B\vert t\vert }-\phi)
\right )^2 .    
\label{fit}
\end{eqnarray}
It has been used for the fit in Fig. 5. Only the very first oscillating term 
in (\ref{solut}) is taken into account in this expression, because other 
terms are more strongly damped with $\vert t\vert $. It is important that the
 experimentally measured values of the diffraction cone slope $B$ and the 
total cross section $\sigma _t$ of the same experiment mostly determine the 
shape of the elastic differential cross section in the Orear region of 
transition from the diffraction peak to large-angle parton scattering. The
value $Z=4\pi B/\sigma _t$ is so close to 1 at 7 TeV that the fit is 
extremely sensitive to $f_{\rho }$ because $\ln (Z/f_{\rho })$ in the first term
determines the slope in this region. Therefore, it becomes possible for the 
first time to estimate the ratio $\rho _l$ outside the diffraction cone directly
from fits of experimental data.

\begin{figure} % 5

\includegraphics[width=\textwidth]{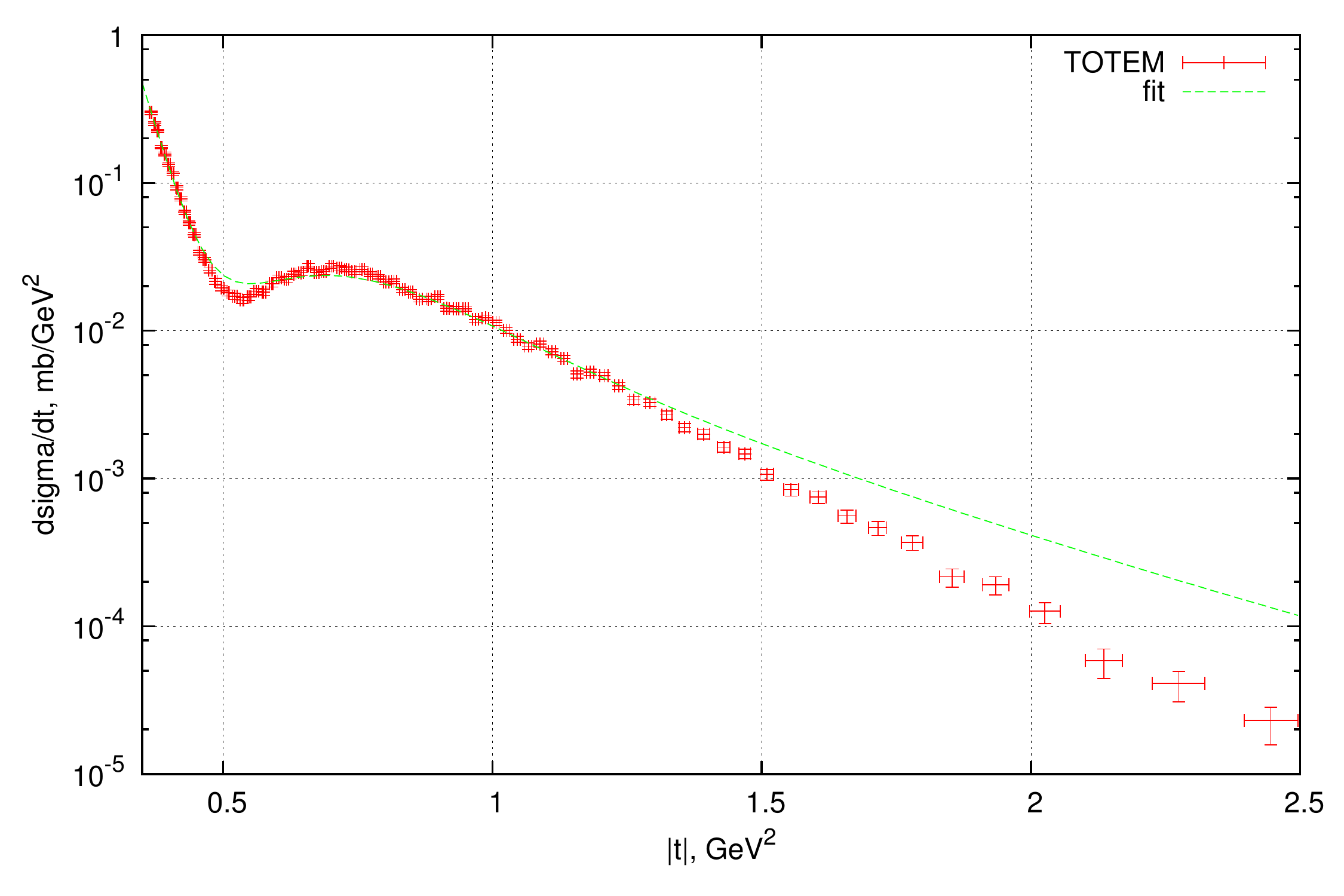}

Fig. 5. The fit of the differential cross section of elastic proton-proton 
scattering at $\sqrt s$=7 TeV in the region beyond the diffraction 
peak according to the predictions of the unitarity condition \cite{dnec}.
Dots - experimental data, line - theoretical approximation.

\end{figure}
                               
Moreover, in footnote 2 in Ref. \cite{anddre1} it was mentioned that 
Eq. (\ref{linear}) is in fact an equation for $\theta ^{1/2}{\rm Im} A(p,\theta )$.
The factor $\theta ^{1/2}$ was omitted in this review and all previous papers
because it was assumed that "retaining it would exceed the accuracy of the
derivation" of the equation. However, it will be worthwhile to take it
into account in the future as well, multiplying the right-hand side of 
(\ref{fit}) by $\vert t\vert ^{-1/4}$. This would slightly improve the fit 
in Fig. 5. 

We note that this shape of differential cross section (\ref{fit}) differs
from formula (\ref{adya}), first of all, because of the suppression of 
oscillations by the exponential factors in front of them, which decrease 
much more strongly than the leading exponent.
In (\ref{adya}), the exponent is common for main and oscillating terms, while
in (\ref{fit}), the oscillations are strongly damped. They may give rise 
to the dip adjusted to the diffraction cone if their amplitude is sufficiently 
large.
The small secondary damped oscillations at larger values of $\vert t\vert $ 
have been seen at comparatively low energies (see Ref. \cite{adg}) but have not 
yet been noticed at the LHC. We stress that the fit (\ref{fit})
contains only three adjustable parameters: the overall normalization $p_1$, 
the amplitude of oscillations $p_2$, which determines the depth
of the dip, and $f_{\rho }$, which helps find the ratio $\rho _l$ outside the
diffraction peak from the slope of the differential cross section there.

\subsubsection{Overlap function and the eikonal}

Both the overlap function and the eikonal are subject to the unitarization 
procedure, albeit in somewhat different approaches. Therefore, it is instructive 
to compare their different forms.
 
We discuss what shapes of the overlap function can be considered as
suitable for further use. One of the assumptions used in solving the 
unitarity equation was the smallness of $F(p,\theta )$ in the Orear region.
The results in \cite{dnec, ads} give strong support to the validity 
of this assumption. The overlap 
function was calculated there directly from experimental data, by subtracting 
the elastic contribution $I_2$ from the left-hand side of the unitarity equation 
without appeal to any model. It is described by the formula:
\begin{eqnarray}
F(p,\theta )=               \
16p ^2\left (\pi \frac {d\sigma }{dt}/(1+\rho ^2)\right )^{1/2}-\nonumber \\
\frac {8p^4f_{\rho }}
{\pi }\int _{-1}^1dz_2\int _{z_1^-}^{z_1^+}dz_1\left [\frac {d\sigma }{dt_1}
\cdot \frac {d\sigma }{dt_2}\right ]^{1/2}K^{-1/2}(z,z_1,z_2).
%\nonumber
\label{overl}
\end{eqnarray}
Here, $z_i=\cos \theta _i; \;\;\; K(z,z_1,z_2)=1-z^2-z_1^2-z_2^2+2zz_1z_2$;
 $\;\;\; z_1^{\pm }=zz_2\pm [(1-z^2)(1-z_2^2)]^{1/2}$.

The result at 7 TeV is shown in Fig. 6.

\begin{figure}

\includegraphics[height=8cm]{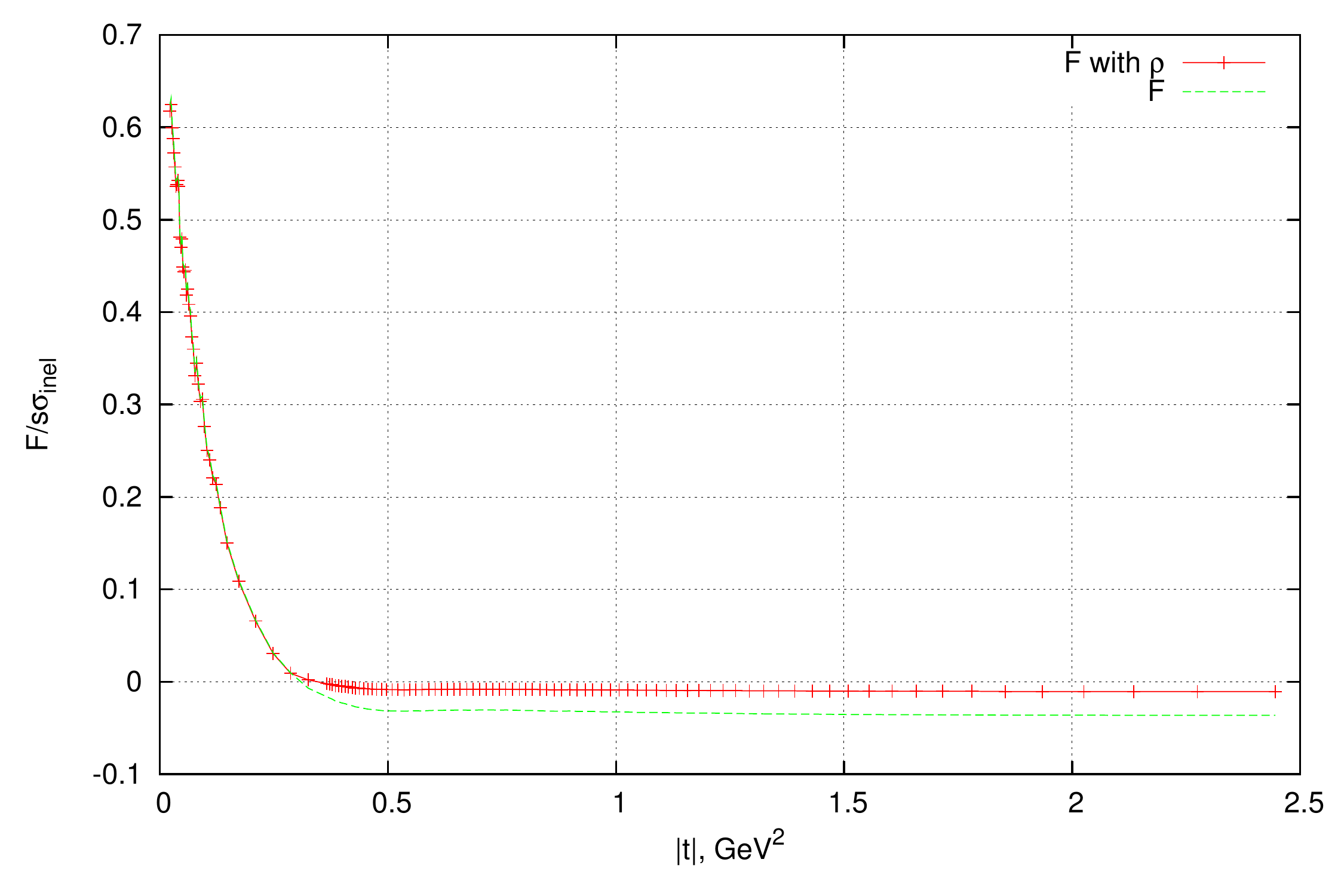}

Fig. 6. The overlap function at $\sqrt s$=7 TeV obtained from the
unitarity condition with substitution of experimental data about the 
differential cross section \cite{dnec}. \\ It is large in the diffraction cone 
and negligibly small outside it. The line nearest to the abscissa axis takes 
into account the real part of the amplitude. The farthest one is computed with
$\rho =0$.

\end{figure} 
Certainly, the shadow of inelastic processes represented by the overlap function
dominates within the diffraction peak. But it is extremely small outside. 
It is even smaller at the LHC energies \cite{dnec} than at lower ones 
\cite{ads}, where a similar behavior of the overlap function at large 
$\vert t\vert $ was observed previously. Hence, this assumption is well founded.

Moreover, it is quite understandable that $F(s,t)$ is very small at large
$\vert t\vert $ in Fig. 6. This shows that its fit by the solution of the 
unitarity relation has been done by the proper eigenfunction (\ref{solut}) 
with the correct eigenvalues of the integral equation. 

It is tempting to solve the nonlinear inhomogeneous unitarity equation 
(\ref{unit}) by iterations. That has been attempted several times 
\cite{hove, amati, cott, anddre1}. The main problem is the choice of the
overlap function. The simplest ansatz is the Gaussian form at all transferred 
momenta. The argument in favor of it is just that it plays the decisive role
in the diffraction cone, where the elastic amplitude has a Gaussian shape.
BBut the results fail to describe the Orear regime. This may be ascribed 
to the role of phases of inelastic processes, that determine the genuine shape 
of the overlap function, or/and to the improper approximation of $\rho $ by a 
constant outside the diffraction cone. Again, similarly to the situation in the 
$b$-representation, the tiny details of the shape prevent from the proper 
outcome. No approximations for the overlap function demonstrated in Fig. 6 
have yet been proposed.

It is instructive to confront the shape of the overlap function $F(s,t)$ with 
results obtained in the impact parameter interpretation of proton-proton 
scattering. They were presented in 
Refs \cite{htk74, as80} for ISR data and are demonstrated in Figs 7, 8. 
The $b$-transformed amplitude $h(s,b)$, the overlap function $F(s,b)$ and the 
eikonal $\Omega (s,b)$ are shown in Fig. 7 at the energy $\sqrt s=52.8$ GeV
\cite{htk74}. The transformed amplitude is almost Gaussian from the center
to 2 fm with little flattening near the center. There is a tail beyond 2 fm
with a much flatter slope. The flattening of the overlap function at the 
center is much stronger, while the eikonal is steeper there.
Hence, one should not identify these three curves at small $b$, even though 
they almost coincide beyond 2 fm. 

\begin{figure}
\includegraphics[width=\textwidth, height=8cm]{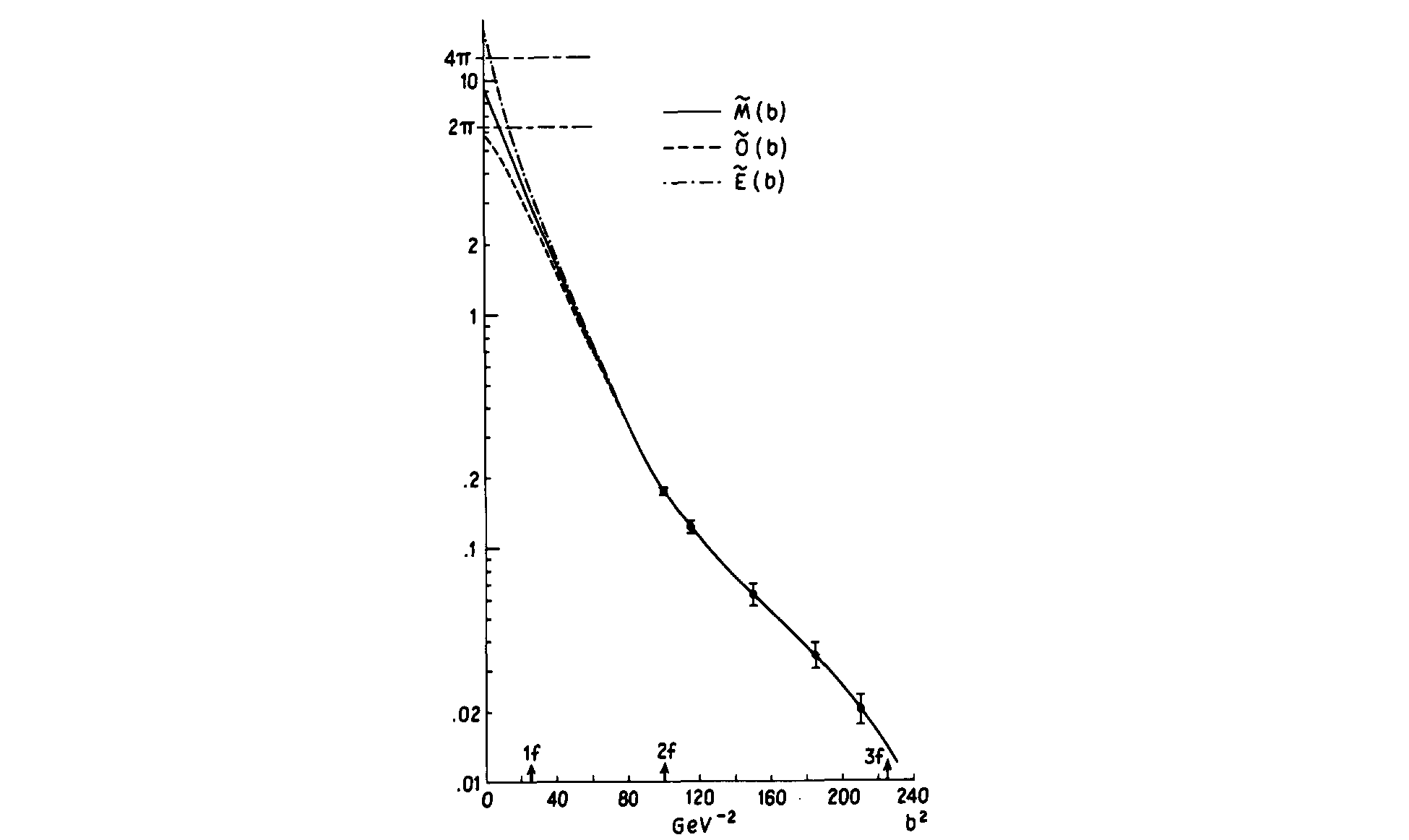}

Fig. 7. The shapes of the amplitude, overlap function and eikonal extracted
from experimental data at $\sqrt s=52.8$ GeV as functions of
impact parameter squared (borrowed from \cite{htk74}). In the notations 
of this review, the amplitude $h(b)=\tilde{M}(b)/8\pi $, the overlap function 
$F(s,b)=\tilde{O}(b)/8\pi $, the eikonal $\Omega (s,b)=\tilde{E}(b)/
8\pi $. The corresponding space scales are shown in the abscissa axis.

\end{figure}

\begin{figure}
\includegraphics[width=\textwidth, height=8cm]{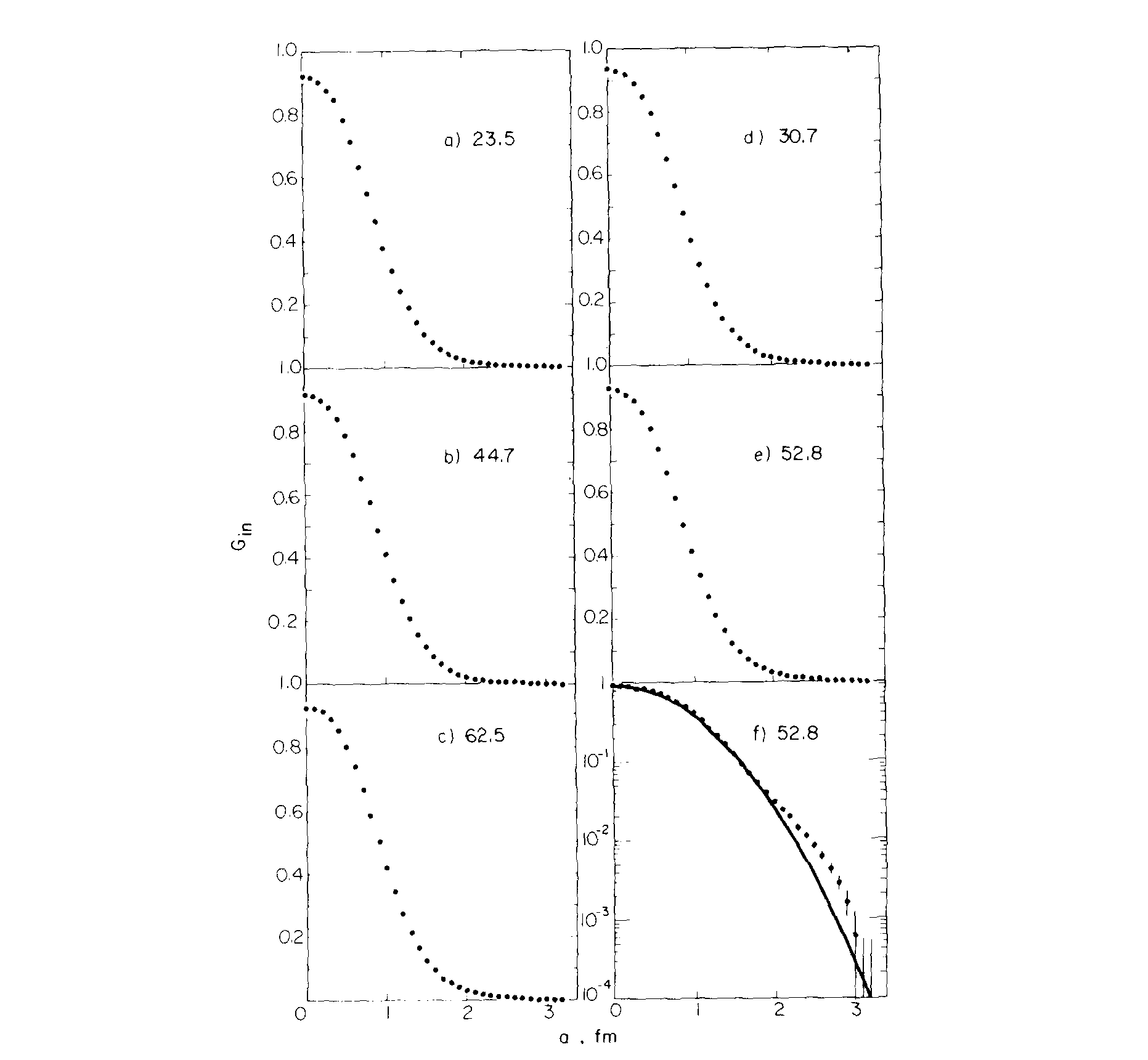}

Fig. 8. The overlap functions at ISR energies as functions of the 
impact parameter look similar (borrowed from \cite{as80}). The solid line at
$\sqrt s=52.8$ GeV is explained in the text.

\end{figure}

Similar features are seen in Fig. 8, taken from
\cite{as80}, where $F(s,b)$ at the same energy is displayed. The solid line
on the logarithmic scale is a Gaussian adjusted to fit at $b=0$ and $b=1.6$ fm.
A Gaussian adjusted between 0.6 and 1.6 fm would be higher at $b=0$ and would
require additional flattening. This flattening at small $b$ corresponds 
directly to negative values of $F(s,t)$ at large $\vert t\vert $ seen in Fig. 5.
In the same way, slight variations of eikonal $\Omega (s,b)$ at small $b$ may
lead to drastic disagreement of model fits with experimental data. Therefore,
their success or failure at large $\vert t\vert $ depends on the accuracy of the
chosen form of the eikonal at low $b$. A long tail above the solid line for 
large impact parameters is clearly seen in Fig. 8f. These Figures demonstrate 
how accurate model formulas must be to correctly reproduce  either the overlap
function or the eikonal if the final goal is to describe the differential cross
sections outside the diffraction peak.

A small "edge" correction to the Gaussian shape of the eikonal has been claimed
to be necessary for fits of experimental data at ISR energies on increasing
total cross sections and structures of the differential cross sections in Refs 
\cite{hv74, h76, hv79}. For example, the correction factor $k$ with some
specific dependence on the impact parameter was introduced \cite{hv79} into the 
overlap function $F(s,b)$. It changes the shape at small $b$ and makes it 
similar  to that shown in Fig. 7:
\begin{equation}
F(s,b)=P\exp (-b^2/4B)k(s,b\exp [-\gamma ^2b^2/4B]).
\label{fsbk}
\end{equation}
It turns out that, in the $t$-representation, the corresponding overlap function 
$F(s,t)$ has two zeros at $\vert t\vert $=0.645 and 3.83 GeV$^2$ and 
becomes practically indistinguishable from zero already at $\vert t\vert >$3.5 
GeV$^2$. The last statement is in full agreement with the conclusions in Refs 
\cite{adg, dnec}.

Although the overlap functions in Fig. 8 look quite similar to each other,
there is a slight difference, which was analyzed in \cite{hv79, as80}.
This difference reveals itself in a small increase at the level of 4$\%$ of the
overlap function, with an energy increase at the impact parameters (radii)
about 1 fm, which implies the peripheral origin of this phenomenon. That was
also discussed earlier \cite{hmv74}. Moreover, in Ref. \cite{as80}
which deals with the direct analysis of experimental data at ISR, a shoulder of
the overlap function at 2.3 fm was noticed. Its origin is unknown.
                                                                        
The increase in the peripheral region about 16$\%$ for $Sp\bar p S$ data is 
reported in the latest review \cite{uama}. Our recent results (to be published)
indicate that it is twice larger at LHC.

The overlap function in the $b$-representation is used in Ref. \cite{trt12}
to distinguish between the mechanisms of absorption and reflection with the
help of the unitarity equation. In the latter case, the differential cross 
section at large momentum transfers is predicted to be 4 times larger.

The impact parameter picture used in almost all phenomenological models
is very helpful for a qualitative description of the process. However,
the forms of the eikonal in the $b$-representation turned out to be very approximate. 
In our opinion, their wide use in most papers dealing with extension to 
larger angles suffers from this deficiency. There are some arguments \cite{cw2}
that the eikonal approximation is only valid for sums of leading terms of 
the tower diagrams, but it is not correct in general. It is applicable to almost 
collinear processes only and does not properly take the separation due to 
transverse momenta into account. That is why the quasi-eikonal models were
developed where the intermediate states take inelastic diffraction processes 
into account, in addition to elastic ones. As a result, formulas like
(\ref{gpps}), (\ref{gpps1}) were proposed. The eikonal does not properly reproduce  
the $s$-channel cuts of the scattering amplitude due to multiple scattering 
\cite{lapo}. By itself, it does not guarantee precise unitarization. Moreover, 
the procedure of unitarity corrections is not well defined, because it can be 
implemented differently. The accuracy of 
unitarity relation (\ref{unib}) in the $b$-representation is also not absolutely
clear, as discussed above, while its use is mandatory for interpretation of
experimental data. That is why the model predictions shown in Fig. 2 fail 
to explain the data. 

There is a drastic difference between the use of the Gaussian shape for the
amplitude in the $s$-channel unitarity condition and the same shape for the
overlap function, as well as its use directly in the $b$-representation. 
The exponential decrease (see Eq. (\ref{diff}))
of the differential cross section in the diffraction cone as a function of
$\vert t\vert $ (or Gaussian for angles) is an experimental observation. It can
be used anywhere within the its applicability range, as it was done, for
example, in solving Eq. (\ref{unit}). Hence, this solution is quite
successful in fits of experimental data in the Orear region. The same shape
cannot be used for the $t$-dependence of the overlap function, although it 
plays an important role in the formation of cone behavior.

It is often argued that the Fourier transform of the Gaussian is a Gaussian
and therefore this shape can also be used in the $b$-representation.
While the first part of the statement is correct, the second is wrong.
The tails of the differential cross sections are very sensitive to small $b$.
Slight variations of this shape at small impact parameters lead to crucial
changes in the behavior of the amplitude at large transferred momenta. Therefore,
the predictions shown in Fig. 2, which use the impact parameter profiles 
close to
Gaussian ones even in the vicinity of $b$=0, are still successful inside the 
diffraction cone  but completely fail outside it, where central collisions 
play an important role. It is very difficult in a particular model to guess the 
proper decline from the Gaussian shape at small impact parameters, which 
drastically influences the differential cross section at large transferred 
momenta.

Therefore, attempts to use the non-Gaussian electromagnetic form factors were of
some help in improving the situation, because they are closer in shape to the
eikonal demonstrated in Figs 7 and 8. Further progress in this 
direction is necessary in order to understand the geometric content of the 
interaction region in ordinary space and time.

Nevertheless, it is hardly justified to blame the phenomenological model
builders for their failure to predict the behavior of the differential 
cross sections at large transferred momenta, where it is many orders 
of magnitude lower than in the diffraction peak. The great and important 
task of fits of the energy behavior of total and elastic cross
sections, ($s, t$)-dependence of the differential cross section, and the
ratio $\rho $ in a wide interval of energies and transferred momenta cannot be
accomplished without free parameters and the physical intuition of model
builders. The switch to higher energies allows eliminating corrections
due to secondary Reggeons and improving the fits. There is hope of
gaining clearer insight into the geometrical picture of hadron interactions.

\subsubsection{Real part of the elastic scattering amplitude at nonzero
transferred momenta}

There are no reasonable arguments to neglect the $t$-dependence of the ratio
$\rho (s,t)$ in (\ref{rho}) or of the phase $\zeta $ in (\ref{zeta}). This 
dependence seems to be important, even inside the diffraction cone, albeit the 
values of $\rho $ are small there. Using formula 
(\ref{rhodit}) and assuming that ${\rm Im}A(s,t)$ determines mainly the shape 
of the differential cross section in this region, we find that the real part
must vanish at
\begin{equation}
t_0=-2\frac {d\ln \sigma _t(s)/d\ln s}{dB(s)/d\ln s}.
\label{zerot}
\end{equation}
With the $\ln ^2s$-dependence of $\sigma _t$ (\ref{cst}) and $B(s)$
(\ref{wid}) and using relation (\ref{sas}), we have
\begin{equation}
\vert t\vert =2/B=16\pi /\sigma _t,
\label{t0sb}
\end{equation}
and hence $t_0\rightarrow 0$ at $\sigma _t\rightarrow \infty $. The estimates at 
LHC energies are $0.1 < \vert t\vert < 0.3$ GeV$^2$. Notably, they agree
with the results obtained in the models in \cite{sel12, kopopo}.

There were several attempts to consider the behavior  of $\rho (s,t)$ at larger
transferred momenta in Refs. \cite{as80, krol, ggk75, sel12, klk87}.
The main efforts were spent on preventing differential cross sections from 
vanishing at those values of $t$ where the imaginary part of
the amplitude is zero in a particular model. The ratio $\rho (t)$ should be
infinite, e.g., as in the models in Refs \cite{as80, sel12}. The number 
of zeros of the imaginary part             
is sometimes greater than one. This is typical in the Fraunhofer diffraction or
in models with electromagnetic form factors. Therefore, the singularities of 
$\rho (t)$ appear
at different $t$ in different models. The real part of the amplitude fills in
these kinks leaving some traces like shoulders or dips in the differential
cross sections. For example, it is predicted in Ref \cite{sel12} that for 
$pp$-scattering at 8 TeV such traces appear at $\vert t\vert \approx 0.35$ GeV$^2$ 
and at 1.5 GeV$^2$. 

In Refs \cite{krol, ggk75}, the dispersion relation between the phase and the 
modulus of the elastic amplitude considered in Refs \cite{odor68, aes71} was 
used with some input for the modulus fitted to the experimental data at  
laboratory energies above 100 GeV. The conclusion was that the real part 
exhibits a zero in the $t$-distribution above 200 GeV, which moves away from 
the forward direction as the energy increases. 

In Ref. \cite{klk87}, the eikonal approximation was used following the proposal
in Ref. \cite{cahn82}. Information about the interference region with
a Coulomb amplitude similar to that in Eq. (\ref{inter}) was inserted into the 
total amplitude, with the result
\begin{eqnarray}
A(s,t)=-\frac {8\pi \alpha }{\vert t\vert }sf_1(\vert t\vert )f_2(\vert t\vert )
e^{i\alpha \Phi }+ \nonumber   \\
is\sigma _te^{Bt/2-i\zeta (s,t)}\left [ 1-i\alpha 
\int_{-\infty }^0dt'\ln \frac {t'}{t}\frac {d}{dt'}\left (
f_1(\vert t'\vert )f_2(\vert t'\vert )e^{Bt'/2-i(\zeta (s,t')-\zeta (s,0))}
\right ) \right ].
\label{klk}
\end{eqnarray}
The $t$-dependence of the phase was parameterized with the help of 5 parameters 
as
\begin{equation}
\zeta (t)=\zeta _0+\zeta _1(t/t_0)^{\kappa }e^{\nu \vert t\vert}+
\zeta _2(t/t_0)^{\lambda }, \;\;\; t_0=-1 {\rm {GeV}}^2.
\label{klkp}
\end{equation}
The results showed that the phase (related to $\rho $ by (\ref{rzet})) 
increases from values close to zero at $t$=0 to about 0.5 in the interval 
$0.5<\vert t\vert <1$ GeV$^2$. This conclusion disagrees with results in
Refs \cite{sel, sel12}, as well as with the arguments presented below.

A more general approach using the $s$-channel unitarity condition was 
developed in Ref \cite{jetp12}. As explained above, the integral 
equation for the elastic amplitude is valid in the Orear region. Its
analytic solution (\ref{solut}) was first obtained in the approximation where 
the values of $\rho $ in $f_{\rho }$ were replaced by their average values in 
the diffraction cone and in the Orear region. No zeros of the imaginary part of
the amplitude were obtained. The dips at 7 TeV and lower energies were 
explained as resulting from damped oscillations. The necessity to introduce 
large negative values of $\rho $ into the Orear region is the main outcome and 
surprise of the fit in Ref. \cite{dnec}. In principle, this could happen if there
were zeros of the imaginary part of the amplitude in this region, which would 
require very large values of $\vert \rho \vert $ near them. But there seem to
be no such zeros there. We discuss this problem in more detail.
  
We first recall asymptotic predictions. It was shown in \cite{mar1} that the 
ratio of the real and imaginary parts of the amplitude can be 
calculated asymptotically at nonzero transferred momenta $t$ as
\begin{equation}
\rho =\rho _0 \left [1+\frac {\tau (df(\tau )/d\tau )}{f(\tau )}
\right ].
\label{rhotau}
\end{equation}
We consider the leading term of solution (\ref{solut}).
With the imaginary part of the amplitude in the Orear region represented as
\begin{equation}
{\rm Im}A_o(s,t)=C_0(s)f(\tau )
\label{ore}
\end{equation}
it is possible to calculate $\rho $.

The very first approximation was to use the first term of the solution 
(\ref{solut}) with average values of $\rho $ both in the diffraction peak 
($\rho _d\approx \rho _0$) and in the Orear region ($\rho _l$) \cite{jetp12}. 
Then the following behavior of $\rho $ was obtained
\begin{equation}
\rho (s,t)=\rho _0\left [1-\frac {a\sqrt {\vert t\vert }}{2}\right ]
\label{rhot}
\end{equation}
where
\begin{equation}
a=\sqrt {2B\ln \frac {Z}{1+\rho _0\rho _l}}.
\label{a}
\end{equation}
We note that $\rho $ passes through zero and changes sign at 
$\vert t\vert =4/a^2\approx 0.1$ GeV$^2$. This agrees with the general 
theorem on the change of sign of the real part of the high-energy scattering 
amplitude, which was first proved in Ref. \cite{mar2}. A similar
effect is discussed in Ref. \cite{sel12}. But it is difficult to obtain 
$\rho _l=-2.1$ as an average of (\ref{rhot}) over the Orear region.

Moreover, this behavior of an unlimited decrease in $\rho $ with $\vert t\vert $
does not look satisfactory. It can in fact be damped if instead of replacing
$\rho $ by $\rho _l$ in the solution, we differentiate $f$ according to 
(\ref{rhotau}), inserting there Eq. (\ref{ore}), i.e., the first term in 
(\ref{solut}). The following differential equation is then obtained
\begin{equation}
\frac {dv}{dx}=-\frac {v}{x}-\frac {2}{x^2}\left (\frac {Ze^{-v^2}-1}{\rho _0^2}
-1\right ).
\label{dvdx}
\end{equation}
Here, $x=\sqrt {2B\vert t\vert }, \; v=\sqrt {\ln (Z/f_{\rho })}$. The 
dependence of $\rho (t)=(Ze^{-v^2}-1)/\rho _0$ can be obtained from Fig. 9.
However, one should read $(1-Z)\rho (t)$ on the ordinate axis in place of
$\rho (t)$. I am sorry for this omission. Unfortunately, the conclusions at
the LHC energies become very indefinite, because $Z$ is very close to zero 
there. The only conclusion is that $\rho (t)$ has a single zero at 
$\vert t\vert \approx 0.3$ GeV$^2$, and it steeply changes
in the Orear region of $0.3<\vert t\vert <1.4$ GeV$^2$.  The result 
shown in Fig. 9 is another extreme approximation compared to Eq.(\ref{rhot}). 

\begin{figure}                                             

\includegraphics[height=8cm]{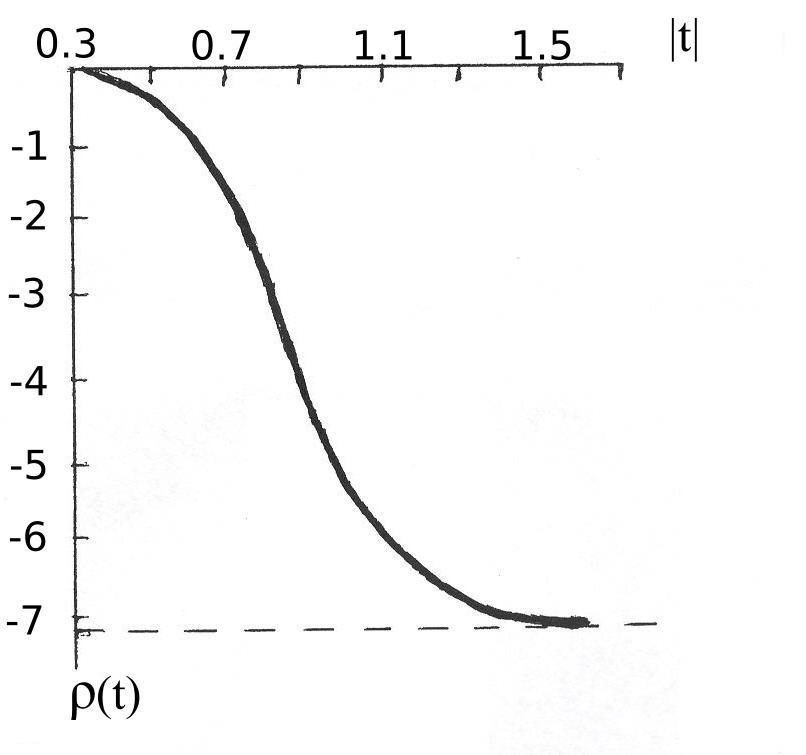}

Fig. 9. The ratio of the real to imaginary part of the amplitude obtained from
the solution of Eq. (\ref{dvdx}) that follows from the unitarity condition 
\cite{jetp12}. The ordinate should be read as $(1-Z)\rho (t)$ in place of
$\rho (t)$. I am sorry for this omission.

\end{figure} 
The bold use of this procedure for derivation of Eq. (\ref{dvdx})
with $\rho (t)$ inserted directly in the solution is, nevertheless, not 
satisfactory, either. The two possibilities above should be considered as two 
extremes for the shapes of $\rho (t)$.

Strictly speaking, the behavior of $\rho (t)$ 
should be taken into account primarely in the integrand. Then, inserting 
expression (\ref{rhotau}) in place of $\rho _l$ in Eq. (\ref{linear}) and 
integrating by parts, we derive the linear integral equation
\begin{equation}
{\rm Im}A(x)=\frac {1}{Z \sqrt {\pi }}\int _{-\infty }
^{+\infty }dy e^{-(x-y)^2} [1+0.5\rho _0^2+\rho _0^2 (y-x)y]{\rm Im}A(y)
\label{nonsymm}
\end{equation}
with $F(p,\theta )=0$ and new variables $x=\sqrt {B/2}p\theta ; \; 
y=\sqrt {B/2}p\theta_1 $.

The kernel of this equation is not symmetric. Its solution has not yet been 
obtained, even numerically. However, some preliminary asymptotic 
estimates can be obtained from it \cite{jetp12}.

In the preasymptotic energy region, we obtained \cite{anddre1} the Orear regime
${\rm Im}A\propto \exp (-ap\theta )\approx \exp (-ap_t)$ with the exponential 
fall-off of the amplitude as a function of angles. We, therefore, seek  
a solution of Eq. (\ref{nonsymm}) in the form ${\rm Im}A(x)=
\exp (-ax\sqrt {2/B} )\phi (x)$. The Gaussian exponential shifts to 
$x-y-a/\sqrt {2B}$.
Replacing it with the $\delta $-function of this argument, we obtain
the equation in finite differences:
\begin{equation}
\phi (x)=Z^{-1}e^{a^2/2B}[1+0.5\rho _0^2(1+\frac {a^2}{B}-ap_t)]
\phi (x-\frac {a}{\sqrt {2B}}).
\label{findif}
\end{equation}
Again, we can not solve it directly, but reach an important conclusion
about the zeros of the imaginary part of the amplitude.
The expression in the square brackets is equal to zero at
\begin{equation}
p_{t0}=\frac {2}{a\rho _0^2}[1+0.5\rho _0^2(1+a^2/B)]\approx 
\frac {2}{a\rho _0^2}.
\label{pt0}
\end{equation}
With the present-day values of $B,\; a,\; \rho _0^2$, this zero would appear at 
extremely large $p_{t0}\approx 20$ GeV. However, zeros of the imaginary part of 
the amplitude in the Orear region just above the diffraction cone might appear 
as zeros of $\phi (x)$ itself. This result does not contradict the above 
statement about the absence of zeros in the case of small oscillatory terms in
the solution of a homogeneous linear integral equation. 

Moreover, the equation tells 
us that $\phi (x)$ and, consequently, the imaginary part of the amplitude can 
have zeros at $x_n=x_0+\frac {a}{\sqrt {2B}}$. On the $p_t$-axis, these 
zeros would be placed at rather short distances from one another. 

In the black disk limit $Z$ tends to 0.5. If $\rho $ loses in the competition 
with $Z$ within $\ln (Z/f_{\rho })$ and the argument of the logarithm becomes 
extremely close
to 1 or even less, that would mean the drastic change of the regime in the 
Orear region \cite{drnp}. What the outcome of the competition between decreasing 
$Z$ and negative values of $\rho $ will be, poses an interesting 
problem. Experimental data at higher energies will be able to give the answer.

As we see, the real part of the amplitude can dominate at large transferred 
momenta according to the unitarity condition. Compared to the imaginary part, 
it can be large and negative there. This conclusion contradicts, for example,
 the results 
of the models in \cite{bour, sel12} with electromagnetic form factors, where the 
dominance of the imaginary part, on the contrary, is claimed everywhere except 
the tiny regions near its zeros placed in the Orear region, in particular. This 
disagrees with the above results. We must remember, however, that $\rho $ is 
infinite at these zeros (see Fig. 10). An analogous behavior of $\rho $ in the 
case of a single zero has been predicted in Ref. \cite{as80} at ISR energies, 
as shown in Fig. 10. 

A similar shape of $\rho $ is 
obtained in Ref. \cite{sel12} at $\vert t\vert \approx $1.5 GeV$^2$, but for 
the energy $\sqrt s$=8 TeV. The real part decreases with $\vert t\vert $.
Therefore, the conclusions in different papers about the behavior of the real
and imaginary parts of the elastic scattering amplitude are contradictory and 
require further theoretical studies and new experimental data.
\begin{figure}
\includegraphics[width=\textwidth, height=8cm]{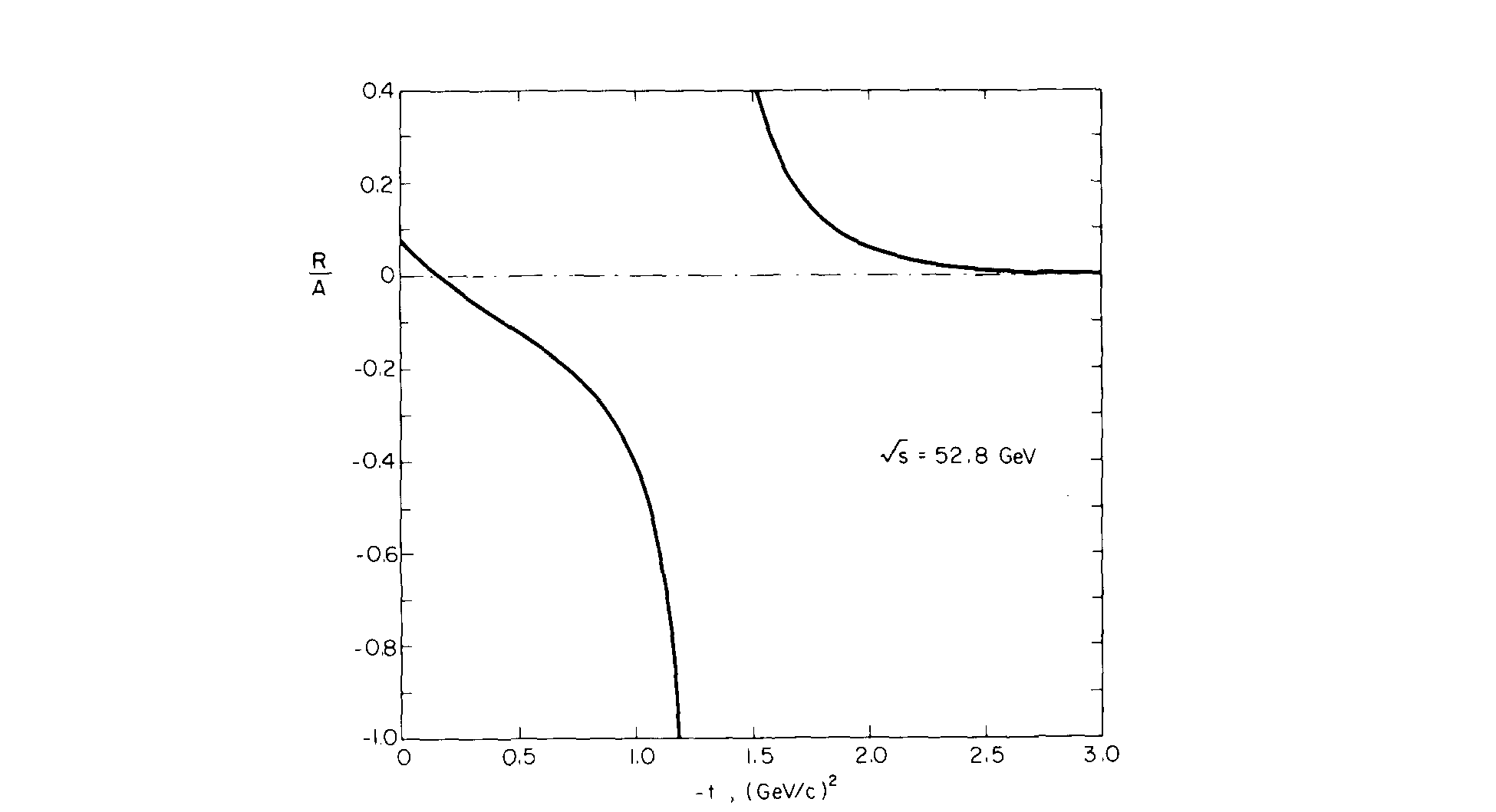}

Fig. 10. The dependence of the ratio of the real part to imaginary part of the 
amplitude (depicted here as $R/A$) on transferred momenta obtained in a 
definite phenomenological fit \cite{as80} of experimental data at 
$\sqrt s=52.8$ GeV. The singularity points out the position where the imaginary 
part is equal to zero.

\end{figure}

\subsection{Scaling laws.}

We have written two formulae (\ref{rhodit}) and (\ref{rhotau}) for
the same function $\rho (s,t)$. Therefore, these two expressions must be 
identical. Equating them, we obtain \cite{drad} the partial 
differential equation
\begin{equation}
p-f(x)q=1+f(x),
\label{partial}
\end{equation}
where $p=\partial u/\partial x ; \; q=\partial u/\partial y; \;
u=\ln {\rm Im}A(s,t); \; f(x)=2\rho (s,0)/\pi ; \; x=\ln s; \; 
y=\ln \vert t\vert $. As usual, the variables $s$ and $\vert t\vert $ should 
be regarded as scaled by the corresponding constant factors $s_0^{-1}$ and
$\vert t_0\vert ^{-1}$.

Eq. (\ref{partial}) can be rewritten as
\begin{equation}
\frac {\partial u}{\partial \ln \sigma _t}-\frac {\partial u}{\partial \ln t}=
1+\frac {d\ln s}{d\ln \sigma _t}.
\label{part1}
\end{equation}

The general solution of Eq. (\ref{part1}) reveals the scaling law
\begin{equation}
\frac {t}{s}{\rm Im}A(s,t)=\phi (t\sigma _t). 
\label{scal1}
\end{equation}

In the asymptotic black-disk limit $\sigma _t\propto \ln ^2s; 
\; \rho (s,0)=\pi /\ln s $, we obtain
\begin{equation}
xp-2q=x+2,
\label{ln2}
\end{equation}
and the solution is
\begin{equation}
u=\varphi _1 (xe^{y/2})+x-y.
\label{soll2}
\end{equation}
This yields the scaling law for
\begin{equation}
\frac {\vert t\vert }{s}{\rm Im}A(s,t)=e^{\varphi _1(\sqrt {\vert t\vert } 
\ln s)}=\phi (z_1),
\label{scal}
\end{equation}
which implies a universal scaling dependence on a single variable $z_1=
\sqrt {\vert t\vert } \ln s$.

We temporarily neglect the contribution of the real part of the 
amplitude to the differential cross section. Then the asymptotic scaling law 
for the differential cross section times $t^2$ should be 
\begin{equation}
t^2 d\sigma /dt=\phi _1^2(\sqrt {\vert t\vert } \ln s).
\label{sl1}
\end{equation}
We note that the additional $t^2$-factor can be replaced by an $s$-dependence if
absorbed in the argument of the scaling function $\phi $. Then this 
formula coincides with that obtained in the geometric scaling approach
\cite{ddd, dddk}. Thus we have proved that the solution of partial 
differential equation (\ref{partial}) with properly chosen $f(x)$
leads to the results known previously about the geometric scaling.

At the same time, Eq. (\ref{partial}) is more general and can be used
for different assumptions about $f(x)$. In particular, the behavior of the 
total cross section at present energies is often approximated by formula 
(\ref{kppp}) as a sum of a large constant term and another term that
increases as some power of energy (see \cite{kopopo} for a recent reference).
In this case, $\rho (s,0)=\pi \Delta /2 $, and the equation is
\begin{equation}
p-\Delta q=1+\Delta.
\label{eDe}
\end{equation}
Its solution is
\begin{equation}
u=\varphi _2(\Delta x+y)+x-y.
\label{solDe}
\end{equation}
From here, we obtain another universal scaling dependence of the differential 
cross section on a single variable $z_2=\vert t\vert s^{\Delta }$ 
\begin{equation}
t^2 d\sigma /dt=\phi _2^2(\vert t\vert s^{\Delta })
\label{sl2}
\end{equation}
which could be valid at preasymptotic energies.

It follows from the expressions described above that the energy dependence of 
the scaling variable is determined by the behavior of the total cross section.
$\vert t\vert \sigma _t$ if only the first term in Eq. (\ref{rhodi}) is used.
Were this scaling valid, one would be able to predict the 
shape of the differential cross section at a higher energy once the total cross
section is known there. The preliminary results of work with experimental data
at energies from ISR to LHC have shown that just this dependence best reproduces 
the similarity of the shapes of the corresponding lines, even though 
their normalization differs somewhat. Further studies are necessary. 

The above scaling laws must be satisfied for the imaginary part of the 
amplitude times the factor $\vert t\vert /s$ (see (\ref{scal})). It follows 
from Eq. (\ref{rhodit}) that the real part satisfies an analogous scaling law
albeit with another factor, which differs in the two cases considered above.
This would lead to the scaling violating terms when the contribution of 
the real part of the amplitude to the differential cross section is taken into 
account. The above scaling dependences of the differential cross section are 
modified as
\begin{equation}
t^2 d\sigma /dt=\phi _1^2(z_1)+0.25\pi ^2\vert t\vert \phi ^{'2}_1(z_1)
\label{slv1}
\end{equation}
and
\begin{equation}
t^2 d\sigma /dt=\phi _2^2(z_2)+0.25\pi ^2\Delta ^2s^{2\Delta }t^2
\phi ^{'2}_2(z_2).
\label{slv2}
\end{equation}
The violation of scaling laws is different in these cases. The first law 
acquires a term with the coefficient depending only on the transferred
momentum, while the second law aquires a term with the coefficient that depends
both on energy and on the transferred momentum.

This violation of scaling laws must be negligible in the diffraction cone
because the squared ratio of the real part to imaginary part -- which is 
crucial for the differential cross section -- is extremely small there. It would 
be interesting to learn about the effect of these terms outside the cone, 
especially in the Orear region of transferred momenta.

We note that at small values of their arguments $z_i$, the scaling functions 
$\phi _i(z_i)$ must be respectively proportional to $z_1^2$ and to $z_2$,
for the differential cross section to be equal to a constant at $t=0$.

Recent fits of TOTEM data have shown \cite{dntot} that the geometric 
$t\sigma _t$-scaling is violated even within the diffraction cone and must
be replaced at present energies by approximate $t^{1.2}\sigma _t$-scaling.

\subsection{Hard scattering at large angles}

\subsubsection{Dimensional counting}

The energy dependence of high-energy scattering processes at a fixed 
center-of-mass angle is of special interest. Dimensional scaling laws have been 
developed as an approach to understanding it. 
The large angle scattering is determined by contributions due to central 
interactions of internal domains inside the colliding particles. The estimates
according to the perturbative QCD become justified due to its asymptotic 
freedom property. They depend on the number of constituent fields of the 
hadrons \cite{matv, brod}. At large $s$ and $t$ and a fixed ratio $s/t$, we have
\begin{equation}
d\sigma /dt\vert _{AB\rightarrow CD}\propto s^{-n+2}f(t/s),
\end{equation} 
where $n$ is the total 
number of fields in $A, B, C, D$ that carry a finite fraction of the momentum.

Assuming the existence of quark constituents, the 
$s\rightarrow \infty $, fixed-$t/s$ prediction for $pp$-scattering 
\cite{brod} is $d\sigma /dt \propto s^{-10}$.

For the elastic amplitude, it is
\begin{equation}
A_1(s,t)\propto \left (\frac {s_0}{s}\right )^{\frac {n}{2}-2}f_1(s/t).
\label{mabr}
\end{equation}
This form can become more complicated for multiple scatterings. For example, 
the lowest order graphs for $m$ rescatterings \cite{land74} behave as
\begin{equation}
A_m(s,t)\propto \left (\frac {s_0}{s}\right )^{\frac {n-m+1}{2}-2}f_m(s/t)
\label{land74}
\end{equation}
and could become the leading ones.
However, due to higher-order corrections, the resulting behavior could change
not so drastically, and the result would be close to the initial estimate 
(\ref{mabr}) as shown in Ref. \cite{mue81}. Further progress beyond the
simple quark counting rules was slowed down by complications in calculating
the enormous number of Feynman diagrams.

\subsubsection{Coherent scattering}

In parallel, there were attempts to explain the 
$\vert t\vert ^{-8}$-regime in $pp$ scattering by dynamical mechanisms 
with the help of simple Feynman graphs. For protons (or their subregions) 
consisting of three valence quarks, we can assume coherent exchange by 
gluons \cite{dl79, dl96, dnaz} or by the color-neutral pairs of gluons 
\cite{ss94} between them. The propagators of three gluons and their couplings
produce an $\alpha _S^6\vert t\vert ^{-6}$-dependence, and
two powers in the denominator are added by kinematical factors. The general 
problem of these approaches is the necessity to introduce 
additional factors in order to preserve
both protons in their initial states in large-angle scattering.
The corresponding powers of the QCD coupling constant should be included,
of course, which leads to possible (strong ?) modifications of the simple
power law. Also, the exchange by three Pomerons instead of the three pairs of 
gluons is possible. Because three colliding quarks share the total energy of 
the proton equally (?), their shares are smaller, and the whole process is 
farther from the asymptotic regime if treated at the parton level. None of 
these questions have been quantitatively resolved yet.

We note that the large-$\vert t\vert $ behavior of Reggeons composed of two
Reggeized partons (quarks, gluons) can be calculated from the BFKL equation
\cite{kwie, kili}.

The multi-Pomeron exchange for hadrons in a state with a minimum number of 
partons was considered in \cite{ka10}. It was concluded that the differential 
cross section factores as a product of two $\vert S_0(s)\vert ^2$
representing the probability of finding the initial and final particles in a "bare"
state and the $d\hat {\sigma}(s,t)/dt$ describing the hard exchange interaction:
\begin{equation}
\frac {d\sigma}{dt}=\vert S_0(s)\vert ^2\frac {d\hat {\sigma}(s,t) }{dt}.
\label{ka10}
\end{equation}
The first factor describes the contribution of large transverse distances,
and the second factor represents the contribution of small ones.  The hard 
exchange is
determined by the Pomeron vertices, which are known semiclassically:
\begin{equation}
\frac {d\hat {\sigma}(s,t) }{dt}\propto g_1^2(t)g_2^2(t)\propto 
(\alpha _S(t))^{\nu}/\vert t\vert ^N
\label{kan10}
\end{equation}
with
\begin{equation}
\nu =n_1+n_2+\vert n_1-n_2\vert, \;\; N=0.5[3(n_1+n_2)+\vert n_1-n_2\vert )-1],
\label{kanc10}
\end{equation}
where $n_i$ are the numbers of valence quarks in colliding hadrons. This leads
to a $\vert t\vert ^{-8}$-behavior for $pp$ and $\vert t\vert ^{-7}$ for $\pi p$.
The quantitative comparison with experimental data is more difficult because of
much smaller values of the differential cross sections in this region and,
correspondingly, larger error bars.

\section{Discussion and conclusions}

The new experimental data of the TOTEM collaboration at the LHC about elastic
scattering of protons at an energy of 7 TeV have revived interest to these processes.
The picture of very short-wavelength hadron collisions has become available,
adding to our insight into the spatial structure of colliding particles and 
providing new intrinsic information pertaining to very short-distance interactions.
The total and elastic cross sections show a stable increase with energy. The 
share of elastic processes increases. The differential cross section has
very intriguing properties. The exponential $\vert t\vert $-decrease persists
at small transferred momenta, analogously to lower-energy data. But the 
diffraction cone slope is larger compared with low energies; it is stable up to 
transferred momenta $\vert t\vert \approx 0.3$ GeV$^2$, then this peak
steepens and a dip appears at $\vert t\vert \approx 0.53$ GeV$^2$, with a
subsequent maximum at $\vert t\vert \approx 0.7$ GeV$^2$. At somewhat larger 
angles, the exponential in $\sqrt {\vert t\vert }$-regime prevails. It is 
replaced by the $\vert t\vert ^{-8}$-behavior at ever larger transferred momenta
$\vert t\vert >2$ GeV$^2$. 
At the same time, we are waiting for measurements at extremely small angles
in the interference region of Coulomb and nuclear amplitudes to gain some
knowledge about the real part of the forward scattering amplitude. 
It would be extremely interesting to learn its energy behavior and check
our predictions from the dispersion relations.

The steeper slopes of the diffraction peak and of the Orear region at higher
energies, and, correspondingly, their smaller extensions clearly demonstrate
that it becomes more and more difficult for a high-energy particle to preserve 
its identity when scattering at larger transverse momenta. 

This increase in the total cross section and, especially, in the 
share of the elastic cross section, as well as the peculiar
change of regimes in the $\vert t\vert $-behavior of the differential
cross section, require a theoretical interpretation. Short of a complete 
theory of hadron dynamics, we have to turn our attention to phenomenological
models and some rare rigorous theoretical relations. The region of large
transferred momenta became an Occam razor for them, as explained above.

The geometric picture of the internal structure of protons and their 
collisions requires  
larger disk radii increasing with energy. Their blackness increases as well. 
Some separate subregions of different sizes and opacity are considered. The 
impact parameter approach is decisive in deciphering this structure. 
At ISR energies, the increase of the total cross section was attributed to some
peripheral regions of nucleons. It is important to juxtapose these findings with
the LHC data.
The approach to the black-disk asymptotic limit has become very interesting. 
The proposal of geometrical scaling reducing the number of independent 
variables is under investigation. At the same time, the scaling law may happen
to be different from the geometrical scaling.

There are many phenomenological models, at our disposal, but it is still
difficult to choose any particular one among them. Most of them are quite successful,
albeit with many adjustable parameters, in describing the energy behavior of
the cross sections and the main bulk of the elastic processes in the 
diffraction cone, but fail in their predictions outside it. The dynamical
origin of many assumptions is still missing. The small details of the
suspected break at small $t$, of the steepened slope and of possible weak 
oscillations over a smooth exponential behavior of the diffraction peak
are under investigation.

There are predictions of several dips and/or visible oscillatory behavior 
imposed on the trend  of a generally  decreasing dependence 
on $\vert t\vert $, which appear at 
larger transferred momenta. As an example, in Fig. 11 borrowed from \cite{isla2} 
the results of some model predictions for the differential cross section of 
proton-proton scattering at $\sqrt s=14$ TeV are shown up to quite high values 
of $\vert t\vert = 10$ GeV$^2$. They differ significantly, and further accurate 
experimental data expected to be obtained in 2015 -- 2016
will surely be decisive in the choice of a model (if any!).
The experience with unsuccessful predictions at 7 TeV in the region 
outside the diffraction cone is not very encouraging.

\begin{figure}                                             

\includegraphics[height=10cm]{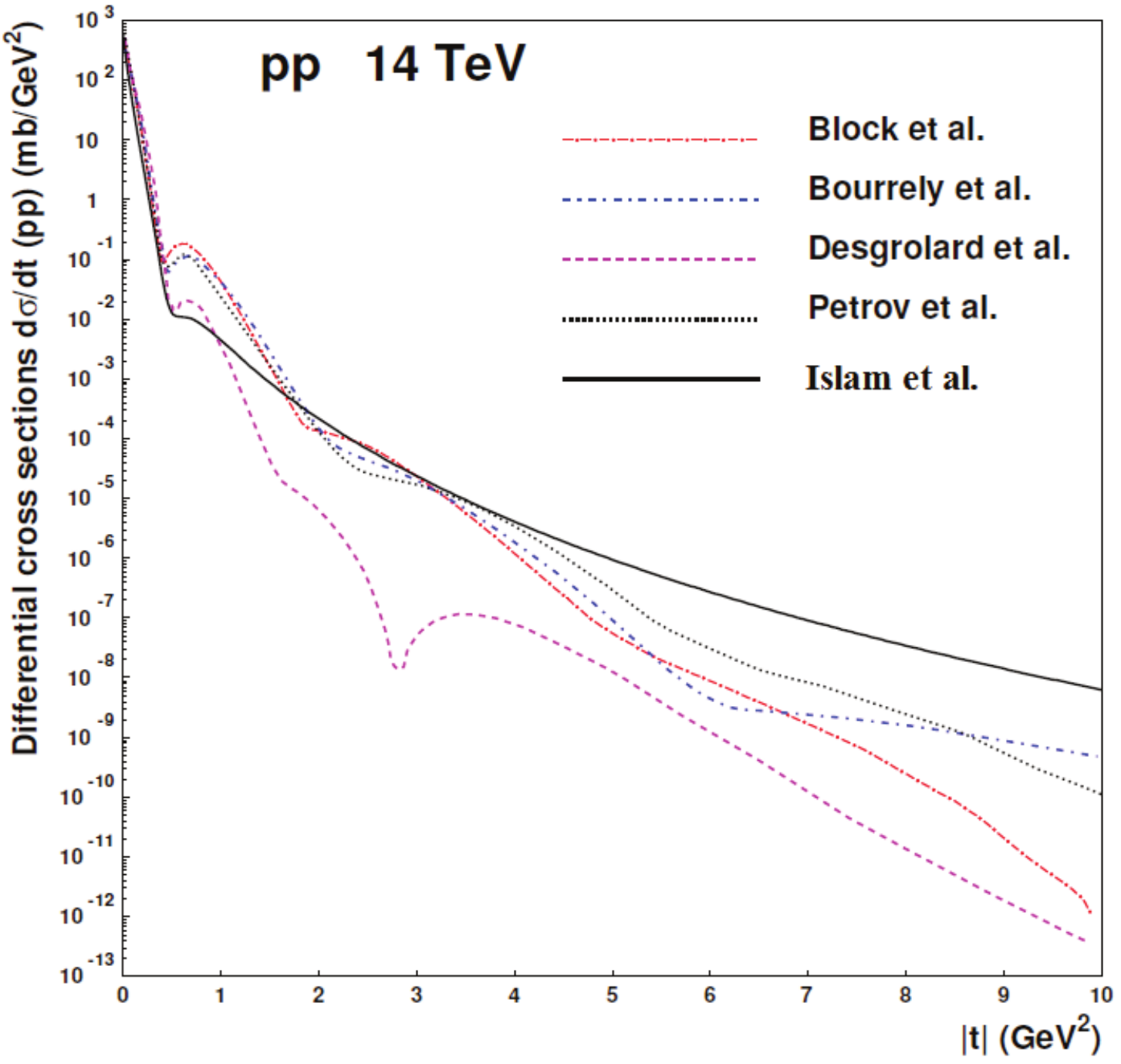}

Fig. 11. Model predictions for the behavior of the differential cross section
of proton-proton scattering at $\sqrt s=14$ TeV presented in Ref. \cite{isla2}.

\end{figure} 

The problem of the behavior of the real part of the elastic 
scattering amplitude at nonforward transferred momenta is becoming very important.
While the imaginary part of the amplitude dominates at small angles in the 
diffraction cone, there are indications that just the real part prevails at 
high transferred momenta. The unitarity condition indicates some ways to
solve this problem. However, there are other approaches with different
conclusions. 

Another important unsolved problem is the behavior of the overlap function. It 
certainly dominates in the diffraction cone, but seems
to become negligibly small outside it. The phases of matrix elements of 
inelastic processes must play an important role in attempts to recover its
shape. However, this presents the extremely difficult theoretical task of 
modeling them. 

Unfortunately, there is little progress in understanding the regime of
power counting for very hard constituent parton scattering, even though some
recent attempts are quite promising.

To conclude, the aforementioned list of problems is not at all complete. 
Many other details should be clarified. Further experimental data will 
definitely shed light on ways to resolve them.

\medskip

{\bf Acknowledgement}

This work was supported by the RFBR grant 12-02-91504-CERN-a and
by the RAN-CERN program.

\end{document}